\documentclass[aps,twocolumn,groupedaddress,superscriptaddress,amsmath,amssymb,prapplied]{revtex4-2}
\usepackage{amsmath}
\usepackage{amssymb}
\usepackage{physics}
\usepackage{dsfont}
\usepackage{graphicx}
\usepackage{dcolumn}
\usepackage{bm}
\usepackage{bbold}
\usepackage[usenames,dvipsnames]{color}
\usepackage[colorlinks,citecolor=Blue,linkcolor=Red, urlcolor=Blue]{hyperref}
\usepackage{tikz}
\usetikzlibrary{decorations.pathmorphing}
\usetikzlibrary{arrows.meta}
\usepackage{braket}
\usepackage[normalem]{ulem}
\usepackage[shortlabels]{enumitem}

\begin{document}

\title{Blueprint of a Molecular Spin Quantum Processor}

\author{A. Chiesa}
\affiliation{Universit\`a di Parma, Dipartimento di Scienze Matematiche, Fisiche e Informatiche, I-43124 Parma, Italy}    
\affiliation{INFN–Sezione di Milano-Bicocca, gruppo collegato di Parma, 43124 Parma, Italy}
\affiliation{UdR Parma, INSTM, I-43124 Parma, Italy}
\author{S. Roca}
\affiliation{Instituto de Nanociencia y Materiales de Arag\'on (INMA), CSIC-Universidad de Zaragoza, 50009 Zaragoza, Spain}
\affiliation{Departamento de Física de la Materia Condensada, Universidad de Zaragoza, 50009 Zaragoza, Spain.}
\author{S. Chicco}
\affiliation{Universit\`a di Parma, Dipartimento di Scienze Matematiche, Fisiche e Informatiche, I-43124 Parma, Italy}    
\affiliation{UdR Parma, INSTM, I-43124 Parma, Italy}
\author{M. C. de Ory}
\affiliation{Centro de Astrobiología (CSIC - INTA), Torrejón de Ardoz, 28850 Madrid, Spain.}
\author{A. G\'omez-Le\'on}
\affiliation{Instituto de F\'isica Fundamental, IFF-CSIC, 28006 Madrid, Spain}
\author{A. Gomez}
\affiliation{Centro de Astrobiología (CSIC - INTA), Torrejón de Ardoz, 28850 Madrid, Spain.}
\author{D. Zueco}
\affiliation{Instituto de Nanociencia y Materiales de Arag\'on (INMA), CSIC-Universidad de Zaragoza, 50009 Zaragoza, Spain}
\affiliation{Departamento de Física de la Materia Condensada, Universidad de Zaragoza, 50009 Zaragoza, Spain.}
\author{F. Luis}
\email{fluis@unizar.es}
\affiliation{Instituto de Nanociencia y Materiales de Arag\'on (INMA), CSIC-Universidad de Zaragoza, 50009 Zaragoza, Spain}
\affiliation{Departamento de Física de la Materia Condensada, Universidad de Zaragoza, 50009 Zaragoza, Spain.}
\author{S. Carretta}
\email{stefano.carretta@unipr.it}
\affiliation{Universit\`a di Parma, Dipartimento di Scienze Matematiche, Fisiche e Informatiche, I-43124 Parma, Italy}    
\affiliation{INFN–Sezione di Milano-Bicocca, gruppo collegato di Parma, 43124 Parma, Italy}
\affiliation{UdR Parma, INSTM, I-43124 Parma, Italy}

\begin{abstract}
The implementation of a universal quantum processor still poses fundamental issues related to error mitigation and correction, which demand to investigate also platforms and computing schemes alternative to the main stream. A possibility is offered by employing multi-level logical units (qudits), naturally provided by molecular spins.
Here we present the blueprint of a Molecular Spin Quantum Processor consisting of single Molecular Nanomagnets, acting as qudits, placed within superconducting resonators adapted to the size and interactions of these molecules to achieve a strong single spin to photon coupling. We show how to implement a universal set of gates in such a platform and to readout the final qudit state. 
Single-qudit unitaries (potentially embedding multiple qubits) are implemented by fast classical drives, while a novel scheme is introduced to obtain two-qubit gates via resonant photon exchange. The latter is compared to the dispersive approach, finding in general a significant improvement. 
The performance of the platform is assessed by realistic numerical simulations of gate sequences, such as Deutsch-Josza and quantum simulation algorithms. The very good results demonstrate the feasibility of the molecular route towards a universal quantum processor. 
\end{abstract}  

\maketitle

\section{Introduction}
\label{sec:intro}
Recent advances in the realization of quantum chips with gradually better performances (both in the number of qubits and in the fidelity of operations) \cite{RevTacchino,Nop2021,HFGoogle,Supremacy,Jurcevic2021,Pogorelov2021,Nation2021,Priviteau2021} could stimulate the question: is it still worth to pursue other routes different from the most established technologies? Are there fundamental issues (not barely technical  problems) which could be more easily tackled by alternative platforms, potentially in the short term?

One should first acknowledge that even the most advanced platforms, i.e. superconducting qubits and ion traps, are still limited by important errors in the manipulation of a relevant number of qubits, thus placing the actual implementation of a universal, error-corrected hardware still far from current capabilities. A promising option to strongly mitigate errors and simplify quantum operations is to move from the binary qubit logic to the use of multi-level logical units called {\it qudits}. In fact, qudits can be exploited to reduce the number of required entangling gates in the synthesis of arbitrary unitaries~\cite{di2015optimal}, with potentially disruptive applications in quantum simulation \cite{TacchinoQudits} and quantum error correction \cite{PRXGirvin,jacsYb,JPCLqec}.  

In this respect, it was shown that molecular spin systems such as Molecular Nanomagnets (MNMs) could constitute the natural playground \cite{Carretta2021}. Indeed, they can provide many low-energy states with naturally long coherence \cite{Zadrozny,Atzori2016,Bader,SIMqubit}, which can be engineered by Chemistry to achieve an impressive degree of control, thus meeting the requirements of tailored quantum computing \cite{Ardavan2015,modules,VO2}, quantum simulation \cite{Lockyer2022,Rogers2022} and quantum error-correction \cite{jacsYb} schemes. The fundamental limitation toward the realization of a MNM-based quantum hardware is represented by the realization of a scalable platform in which individual molecules are initialized, manipulated and read-out. The natural way of achieving this is to export standard methods from superconducting qubits technology \cite{Blais2004,Wallraff2004,Majer2007,Schoelkopf2008}, i.e. to couple MNMs to on-chip superconducting resonators and exploit the tools of circuit quantum electro-dynamics to 'wire them up'. This idea was already put forward in \cite{Jenkins2016}, but it has remained a long-term vision until recent progresses both in increasing the coupling of spins to superconducting resonators \cite{Gimeno2020,Ranjan2020,Rollano2022,Bonnizzoni2023} and in understanding the crucial advantage of molecular qudits for quantum information processing \cite{Carretta2021}. Indeed, on the one hand, the fabrication of nanoscopic constrictions 
on the transmission lines allow one to concentrate the magnetic field in regions where MNMs of the same size can be naturally accommodated. This enhances the spin-photon coupling by orders of magnitude, as experimentally demonstrated \cite{Gimeno2020,Ranjan2020}. On the other hand, the potentiality of properly designed molecular spins to encode qudits with suppressed decoherence \cite{Chiesa2022,Chizzini2022} or embedded quantum error correction \cite{JPCLqec,ErCeEr,npjQI,ChizziniPCCP,Chiesa2022} has been  investigated. 
Hence, we are now in a position to understand and quantify advantages and limitations of a quantum hardware consisting of individual MNMs coupled to superconducting resonators. This requires a scheme for implementing qudit gates and for reading the output, a suitable design of the platform and accurate simulations to figure out where we are and what we can eventually accomplish.

To this end, we present hereafter the blueprint of a Molecular Spin Quantum Processor (MSQP). Both superconducting resonators and MNM qudits embedded within them are designed to achieve a strong coupling to single photons, by locally concentrating the microwave magnetic field and choosing suitable molecular states. \\
We illustrate how to implement a universal set of gates in such a platform and how to readout the final qudit state. In particular, we show that several qubits can be encoded within an individual qudit and single-qudit unitaries corresponding to multi-qubit algorithms can 
be implemented by classical drives sent through proper auxiliary control lines \cite{Jenkins2016}. Moreover, we propose a novel scheme to obtain two-qudit gates by means of a relatively fast resonant real photon-exchange \cite{Carretta2013,SciRep15}, using the tunability of the resonator frequency as the unique knob \cite{Palacios-Laloy2008}. This scheme is compared with a dispersive approach \cite{Gomez2022} for the implementation of two-qudit gates by virtual photon exchange, finding an important speedup for the former and hence an improved fidelity, due to the limited effect of decoherence. The overall performance of the proposal is assessed through realistic simulations of elementary gates and of more complex sequences, such as those involved in the Deutsch-Josza algorithm implemented on a single qudit and in the quantum simulation of interesting physical models. Our simulations take into account the most important errors occurring on a real platform, namely photon-loss and pure dephasing of  individual spins. The very promising results achieved with experimentally reachable conditions demonstrate that a MSQP could be a valid path toward a universal quantum hardware.

The paper is organized as follows: we first present the working principles of the MSQP (Sec. \ref{sec:wp}) and then move to the design of the resonator, along with the estimate of feasible spin-photon coupling strengths (Sec. \ref{sec:layout}). Then we simulate the related performance of the hardware (Sec. \ref{sec:simulations}) for elementary gates and for more complex quantum simulation sequences. We finally summarize results and discuss perspectives for further steps in Sec. \ref{sec:discussion}.

\section{Working principles of the Molecular Spin Quantum Processor}
\label{sec:wp}
We illustrate below the working principles of our MSQC.
In particular, we first introduce the static system Hamiltonian and basic ideas to choose the optimal spin system. Then, we show how each of the five DiVincenzo's criteria can be fulfilled.  \\

For sake of clarity, we start here by considering an elementary unit of the MSQP, consisting of a single resonator embedding two molecular spin qudits, both strongly-coupled to the photon field. The scheme can then be easily generalized to an array of such units, as explained below (Sec. \ref{subsec:scala}). 
Each unit of the scalable architecture is described by the following Hamiltonian:
\begin{equation}
    H = H_p + H_S + H_{Sp}.
    \label{eq:HamTot}
\end{equation}
Here the first term $H_p = \hbar \omega_r(t) \left( a^\dagger a +1/2 \right)$ represents the photon field of tunable frequency $\omega_r(t)=\omega_0 + \delta(t)$ around a central value $\omega_0$, with $a^\dagger$ ($a$) boson creation (annihilation) operators satisfying $[a,a^\dagger]=1$. \\
The second term in Eq. \eqref{eq:HamTot} is the spin Hamiltonian of two spin $S$ qudits:
\begin{equation}
    H_S = \mu_B B \sum_{i=1,2} g_i S_{zi} +  \sum_{i=1,2} D_i S_{zi}^2 ,
    \label{eq:HamS}
\end{equation}
consisting of the Zeeman interaction with an external field and of a zero-field splitting anisotropy term which is essential to make all the energy gaps inequivalent and hence individually addressable by resonant pulses. The eigenstates of $H_S$ are the same of $S_z$ ($\ket{m_i}$), with energies $E_{m_i} = D_i m_i^2 + g_i \mu_B B m_i$ and allowed $\ket{m_i} \rightarrow \ket{m_i \pm 1}$ dipole transitions. [We omit here for simplicity other possible anisotropic terms which do not qualitatively alter our results.]\\
Finally, $H_{Sp}$ represents the spin-photon interaction contribution, given by
\begin{eqnarray} 
    H_{Sp} &=& \sum_{i=1,2} 2 G_i (a + a^\dagger) S_{xi} \\ \nonumber
    &\approx& \sum_{i,m} G_i^m (a \ket{m+1}\bra{m} + a^\dagger \ket{m}\bra{m+1}),
    \label{eq:HamSp}
\end{eqnarray}
where in the second line $G_i^m = G_i \sqrt{S(S+1)-m(m+1)}$ and we have applied the rotating-wave approximation (RWA), which is practically exact in the examined range of parameters. 
Even though not employed in our numerical simulations, the RWA allows us to perform all of the following reasoning with only energy conserving terms.
From Eq. \eqref{eq:HamSp} we note that the coupling of the resonator field with each spin transition $G_i^m$ (and hence the time required to implement two-qudit gates) can be enchanced by proper choice of the spin system and of the transitions, besides resonator engineering. Such a degree of freedom is a rather unique opportunity offered by Molecular Nanomagnets, which can be exploited to significantly speedup two-qudit gates. 
In particular, for a spin $S$ system described by Hamiltonian \eqref{eq:HamS}, transitions between $\ket{m_i=0}$ and $\ket{m_i=\pm 1}$ states (the lowest energy ones in presence of easy plane anisotropy and at low field) are enhanced by a factor $\sqrt{S(S+1)}$ compared to a spin 1/2.
To fully exploit this degree of freedom, we consider in the following two $S=10$ qudits, with $D_i/2 \pi\approx 7-8$ GHz and spin transitions always involving a $\ket{m_i=0}$ state (here and in the following, parameters are in units of $\hbar$).  
Although not corresponding to a specific molecule, the assumed parameters are perfectly reasonable. Indeed, the employed $|D|$ are exactly in between those of the two most famous 
$S=10$ single-molecule magnets, namely Fe$_8$ \cite{Caciuffo1998} and Mn$_{12}$ \cite{Barra1997,Chiesa2017}. Moreover, multi-spin clusters with easy plane anisotropy ($D_i>0$) have also been already synthesized (see, e.g., an $S=6$ Cr$_{12}$ \cite{Collison2003} with $D/2\pi=2.6$ GHz), and examples exist even with much larger $S$ \cite{Waldmann2008}, although with smaller or less characterized anisotropy.\\
In the following subsections we address one by one the  DiVincenzo's criteria.

\begin{figure}[t!]
    \centering
    \includegraphics[width=0.5\textwidth]{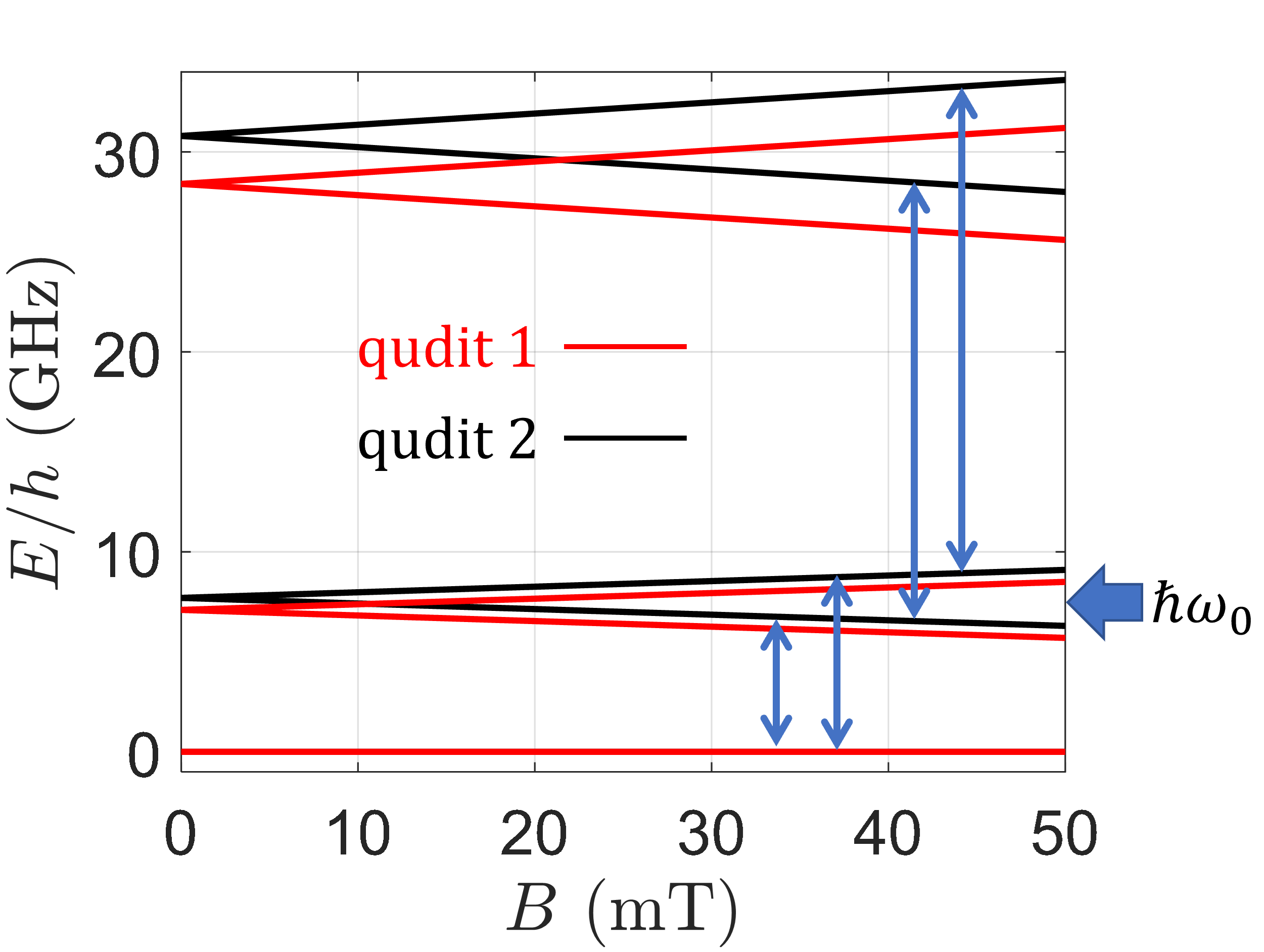}
    \caption{Lowest energy levels for the two qudits used in the reported simulations, with $D_1/2\pi=7.1$ GHz, $D_2/2\pi=7.7$ GHz, $g_i=2$. Blue double-arrows indicate allowed EPR transitions for qudit 2 between $\ket{m_2}\rightarrow \ket{m_2 \pm 1}$ levels, induced by resonant (low-power) microwave classical pulses in the range $5-20$ GHz. An analogous situation is found for qudit 1 (not shown for clarity). The thick arrow on the right indicates the bare resonator frequency $\hbar \omega_0$.}
    \label{fig:levels}
\end{figure}

\subsection{Encoding and Initialization}
In the idle configuration [i.e. with $\delta(t) = 0$], the couplings $G_i^m$ are much smaller than the difference between the bare photon ($\hbar \omega_0$) and spin [$E_{m m^\prime}^{q_i} = E_{m}^{(q_i)}-E_{m^\prime}^{(q_i)}$] excitation energies, for each qudit $q_i$. Hence, the eigenstates of Hamiltonian \eqref{eq:HamTot} are practically factorized and can be expressed as  $\ket{m_1 m_2} \otimes \ket{n_p}$, with  $S_{zi} \ket{m_i} = m_i \ket{m_i}$ and $a^\dagger a \ket{n_p} = n_p \ket{n_p}$.  \\
Then, we encode two-qudit logical states in the zero-photon subspace $\langle a^\dagger a \rangle = 0$, i.e. we represent the two-qudit wavefunction as $\ket{pq} \equiv \ket{m_1 m_2} \otimes \ket{n_p=0}$. This choice reduces the harmful effect of photon loss. Hereafter, we re-label the states with $p,q = 0,...,2S$ in order of increasing energy. With the parameters considered in the text (i.e. low $B$ and $D_i>0$), the sequence of single-qudit states is given by $m_i = 0,-1,+1,-2,+2,...$, as shown in Fig. \ref{fig:levels}. \\
By working at a base temperature below 10 mK (corresponding to a frequency of $\approx 0.2$ GHz), high-fidelity initialization of each qudit in its ground state can be achieved by cooling.

\subsection{Single-qudit gates}
\label{subsec:1qudit}
In analogy to what is usually done for qubits, a universal qudit set can be obtained by combining general single-qudit gates with one two-qudit entangling gate \cite{Brennen2005,Chizzini2022}. \\
As far as the former are concerned, any single-qudit unitary can be decomposed into $\sim d^2$ Givens rotations \cite{Brennen2005} between consecutive $\ket{m} \rightarrow \ket{m\pm 1}$ transitions.  \\
These transitions are indicated by blue arrows in Fig. \ref{fig:levels} and can be accomplished by classical control drives $H_1(t) = B_1 \theta(|t-t_0|-\tau) \mu_B \cos(\omega t +\phi) \left(  g_1 S_{y1} + g_2 S_{y2} \right)$ resonant with the addressed pair of levels. With the here employed parameters, this sets the driving frequency in the $5-20$ GHz range, perfectly achievable using low-power pulses, i.e. $B_1 \sim 1-5$ G. Using 
larger values of $B_1$ (faster pulses) would only increase leakage towards other transitions (different from the addressed one) without significantly reducing the time to implement a generic algorithm. Indeed, even with small $B_1$, the latter is largely dominated by the duration of two-qudit gates, as discussed below. 
In the near future, the performance of single qudit gates (both in terms of gating time and leakage reduction) could be further improved by optimal quantum control techniques \cite{Castro2022}.

In the current development stage, noise is particularly important for multi-qubit gates, while manipulations on individual objects (i.e. a single qubit/qudit) are much easier. Hence, an interesting perspective in the near term is to embed several qubits into a single qudit and then re-write multi-qubit algorithms in terms of single-qudit transitions and gates. This requires to properly develop a dictionary to map multi-qubit into single qudit operations. 
Such a dictionary must include the correspondence between the eigenstates of the multi-qubit an of the qudit system. This allocation of levels can be chosen arbitrarily depending on the connectivity between levels in our qudit. For example, we can assign the multi-qubit levels on which most operations are to be acted upon to the levels which can be manipulated faster. Once such a structure is constructed, we decompose the unitary that comprises the whole multi-qubit circuit into single qudit operations (or Givens rotations) using a QR decomposition method \cite{Brennen2005, Castro2022,Chizzini2022}. With this method, we can decompose any unitary matrix $U$ into the product $U = QR$, where $Q$ is an orthogonal matrix that will be the sequence of different pulses and $R$ is a diagonal matrix that will account for the relative phases induced by the operations between the qudit levels. More explicitly, the operations that we apply to our molecule have the following Givens rotation representation $R_{m, m+1}(\theta, \varphi) = g_{m, m+1}(\theta, \varphi)\oplus I_{(\overline{m, m+1})}$, where $g_{m, m+1}(\theta, \varphi) = \left[\begin{array}{cc} c & s \\ -s^* & c \end{array}\right]$, with $c = \cos(\theta/2)$ and $s = -i\sin(\theta/2)e^{i\varphi}$, is the two-level rotation on the levels $m$, $m+1$ and $I_{(\overline{m, m+1})}$ is the identity over the rest of qudit levels. By choosing properly $\theta$ and $\phi$ we can transform our unitary $U$ into the diagonal matrix $R$: $R = \left(\prod_i \prod_m R_{m, m+1}(\theta^i, \varphi^i)\right)^\dagger U$. This diagonal matrix can be further transformed into the identity applying two-level rotations around the Z-axis, which can be achieved from the previous ones: $R^Z_{m, m+1}(\theta) = R_{m, m+1}(\pi/2, \pi)R_{m, m+1}(\theta, \pi/2)R_{m, m+1}(\pi/2, 0)$ [Cf. App. A in \cite{Castro2022} for a detailed depiction of the algorithm]. With these definitions we have everything that is needed to decompose any multi-qubit circuit into a single qudit one. We now need to find the relation between the decomposition parameters ($\theta, \varphi$) and the physical ones that we are able to implement.

In order to get the physical parameters that are needed for implementing such decomposition, we re-write $H_1(t)$ within the RWA and in interaction picture, setting for each  $\ket{m} \rightarrow \ket{m^\prime}$ transition $\hbar \omega =E_{m m^\prime}^{q_i}$. Then we get
\begin{equation}
    H_1^{\rm int} \approx \frac{1}{2} g \mu_B B_1 |\bra{m} S_y \ket{m^\prime}| \left(e^{-i\varphi} \ket{m} \bra{m^\prime}\ + {\rm h.c.} \right)
\label{eq:HamIntDrive}
\end{equation}
Here, the angle $\varphi = \phi - \delta$ results from the phase of the driving pulse, $\phi$, and from the possible phase induced in the transition between levels (here $\delta = \pi/2$, since we have chosen an oscillating field along $y$). 
 From here, the unitary evolution of the system in the interaction picture reads
 \begin{equation}
     U^{\rm int}(t) = {\rm exp}\left[-i \frac{\theta(t)}{2} \left(\cos \varphi \sigma_x^{mm^\prime} - \sin \varphi \sigma_y^{mm^\prime} \right)\right]
     \label{eq:IntUnitary}
 \end{equation}
 with $\theta(t) = g \mu_B B_1 |\bra{m} S_y \ket{m^\prime}| t$ and where the $\sigma_\alpha^{mm^\prime}$ ($\alpha = x, y$) operators act as the Pauli operators in the subspace of the two levels involved in the transition $|m\rangle \rightarrow |m^\prime\rangle$. $U^{\rm int}(t)$ is precisely the kind of rotations we get from our decomposition algorithm: $\theta$ modulates the population transfer between levels and $\varphi$ the relative phase acquired. Therefore, we can construct directly the experimental pulses that are needed to implement any multi-qubit circuit in our molecule from the sequence of operations that we obtain from the QR decomposition.

\subsection{Two-qudit gates}
\label{subsec:2qubits}
Terrific advances have been recently done in increasing single spin to photon couplings~\cite{Gimeno2020,Ranjan2020} and estimates for $G_i/2\pi$ can reach values in the order of $10^2$ kHz for a spin 1/2, as demonstrated in Sec.~\ref{sec:layout} below. Nevertheless, these numbers still require a careful engineering of photon-mediated two-qudit gates in order to keep gating times ($\tau_{2q}$) significantly shorter than the system coherence. Indeed, the established scheme to implement two-qubit gates in the dispersive regime~\cite{Zueco2009}
implies rather long $\tau_{2q}$, thus making its realization practically difficult. Even for qudits, where additional strategies to optimize the coupling and reduce the required interaction time can be considered~\cite{Gomez2022}. To overcome this limitation, we introduce below a novel scheme which significantly shortens $\tau_{2q}$ by operating in the resonant regime. \\
Before illustrating our proposal, we recall the working principles of the dispersive approach.
The idea is analogous to the first proposals on transmon qubits \cite{Blais2004}. If we consider a pair of spins coupled to the same resonator and such that the spin gaps ($E_{m m^\prime}^{q_i}$) are significantly detuned from $\hbar \omega_r$ (i.e. $G_i^m \ll |\hbar\omega_r-E_{m m^\prime}^{q_i}|$), we can derive a second order effective spin-spin interaction mediated by the virtual exchange of a photon \cite{Carretta2021}, of the form
\begin{equation}
H_{\text{eff}} = \sum_{m,m^\prime} \Gamma_{m,m\pm 1}^{m^\prime, m^\prime \mp 1} \ket{m,m^\prime} \bra{m\pm 1,m^\prime \mp 1}
    \label{eq:dispcoupl}
\end{equation}
with $\Gamma_{m,m\pm 1}^{m^\prime, m^\prime \mp 1} \approx G_{1}^m G_{2}^{m^\prime} / \Delta_i$ and $\Delta_i = \hbar \omega_r -E_{m, m \pm 1}^{q_i}$. Now, if the two spin gaps are significantly different $|E_{m, m \pm 1}^{q_1}-E_{m^\prime, m^\prime \mp 1}^{q_2}|\gg \Gamma_{m,m\pm 1}^{m^\prime, m^\prime \mp 1}$, $H_{\text{eff}}$ is ineffective and the two qudits are decoupled. This occurs, for instance, if the two molecules have slightly different $g_i$ or $D_i$  in Eq. \eqref{eq:HamS}. To turn on the mutual spin-spin interaction, we can apply local magnetic fields \cite{Jenkins2016} to make $E_{m, m \pm 1}^{q_1} = E_{m^\prime, m^\prime \mp 1}^{q_2}$, thus activating an oscillation between $\ket{m, m^\prime}$ and $\ket{m \pm 1, m^\prime \mp 1}$ two-qudit states. \\
In the case of qubits, the resulting evolution $U_{XY}(\tau)$ is equivalent to that 
induced by a spin Hamiltonian of the form $H_{XY} = s_{x1} s_{x2} + s_{y1} s_{y2}$, i.e. $U_{XY}(\tau) = {\rm exp}\left[ -i H_{XY} \tau /\hbar \right] $. Hence, the unitary gate $U_{XY}(\tau)$ can be naturally exploited in the quantum simulation of several models which can be mapped to this Hamiltonian, as illustrated below. 
For proper choice of the interaction time $\tau$, this evolution implements entangling gates such as the $i$SWAP or the $\sqrt{i\rm SWAP}$.
The resulting gating time (to implement, e.g., an $i$SWAP) is given by $\tau_{2q} = \pi \Delta/2 (G_i^m)^2$. With the parameters employed here, $\tau_{2q} \sim 6 \,\mu$s, thus making two-qudit gates implemented with the dispersive approach rather slow and hence prone to decoherence.

\begin{figure}[t!]
    \centering
    \includegraphics[width=0.48\textwidth]{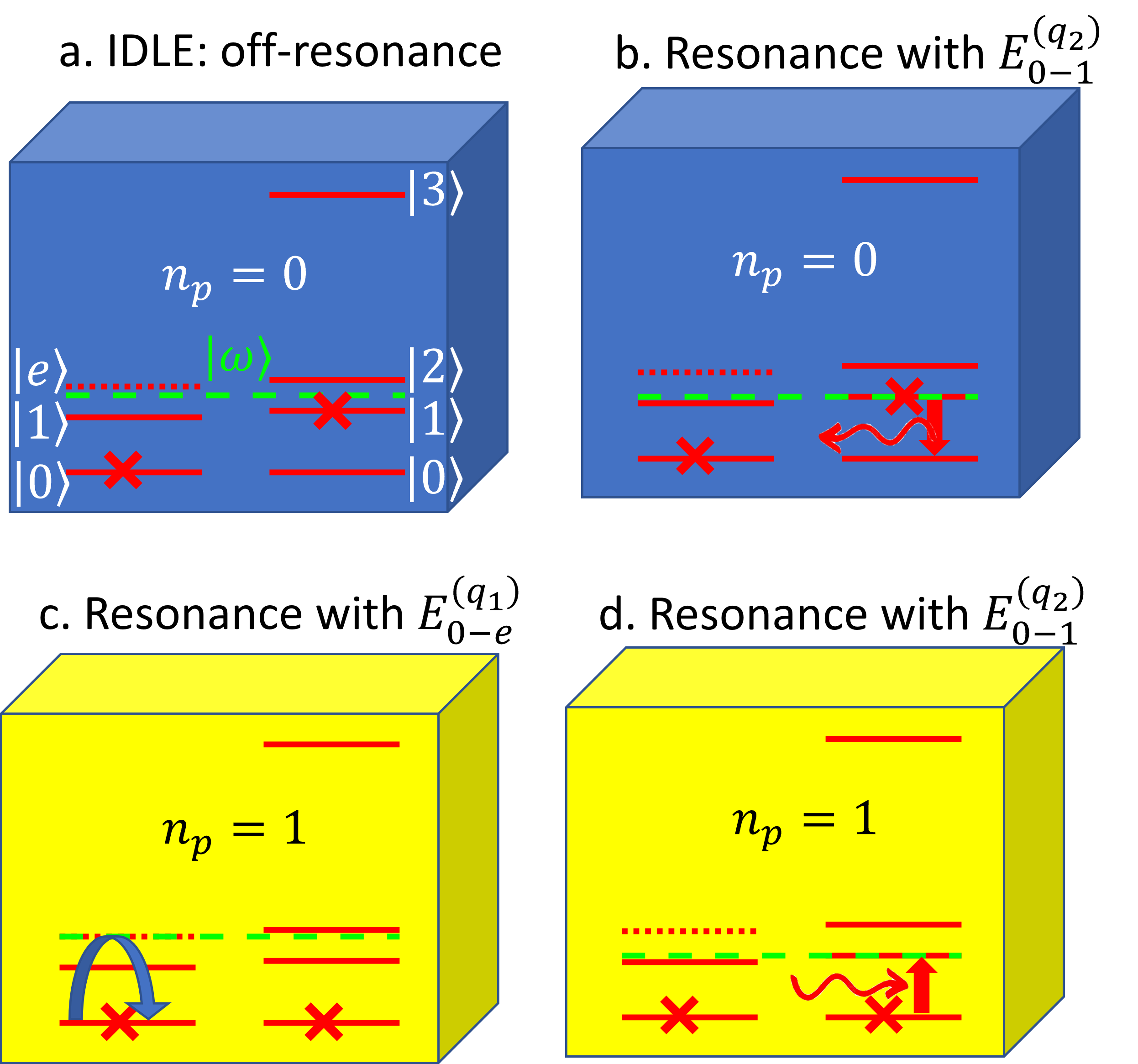}
    \caption{Scheme of the two-qudit resonant gate: Dark (bright) boxes indicate empty (single photon) resonators. Two photon states (included in the simulations) are never populated. Energy levels of the two qudits are indicated by red lines. In this example, we use two levels to encode the logical state of $q_1$ (left) and four to encode $q_2$ (right), while the second excited state of $q_1$ (dotted line, $\ket{e}$) is an auxiliary one exploited during the gate. The dashed green line indicates the resonator frequency which is varied at each step. We illustrate the steps of the gate for a system prepared in $\ket{01}$ (crosses).
    a) Idle state, with no photons (dark) and the resonator out-of-resonance with respect to all qudit transitions. b) The resonator is brought into resonance with $E_{0 1}^{(q_2)}$ and a photon is emitted if $q_2$ was prepared in $\ket{1}$. c) A photon is now in the resonator (bright) and can be absorbed-emitted if $q_1$ was prepared in $\ket{0}$, once the resonator is brought in resonance with $E_{0 e}^{(q_1)}$. d) The resonator is brought again into resonance with $E_{0 1}^{(q_2)}$ and the photon is absorbed, ending up again with $\ket{01}$ and an additional phase. }
    \label{fig:scheme_2q}
\end{figure}

We now move to our novel proposal for two-qudit gates in the resonant regime, i.e. by real (rather than virtual) photon exchange. This leads to gate times scaling linearly (instead of quadratically) with $G_i^m$, achieving a significant speedup compared to the dispersive regime. In particular, we consider a {\it qudit controlled-phase gate}, in which a desired phase $\varphi$ is added only to a specific component $\ket{\bar{p}\bar{q}}$ of the two-qudit wave-function. This corresponds to the unitary transformation  
\begin{eqnarray} \nonumber
    U_\varphi^{\bar{p}\bar{q}} &=& e^{-i \varphi} \ket{\bar{p}\bar{q}} \bra{\bar{p}\bar{q}} \\ 
    &+& \sum_{p,q} \ket{pq}\bra{pq} (1-\delta_{pq,\bar{p}\bar{q}}).
    \label{eq:cphi}
\end{eqnarray}
To minimize the duration of the two-qudit gate, we exploit only transitions among the lowest $m$ states (i.e. $\ket{m=0} \leftrightarrow \ket{m=\pm 1}$) in the level diagram of Fig. \ref{fig:levels}, thus getting an enhancement of the coupling of $\sqrt{S(S+1)} \approx 10$. In the example below, this allows us to directly implement $U_\varphi^{01}$, which can be easily transformed into any other $U_\varphi^{pq}$ gate by adding single-qudit gates implemented via much faster classical resonant pulses. This not only guarantees qudit universality, but also a relevant flexibility in the two-qudit gate, which could be very useful to reduce the depth of many algorithms. \\
To illustrate the implementation of a generic $U_\varphi^{pq}$ gate, we refer to the scheme reported in Fig. \ref{fig:scheme_2q}. In particular, we consider a three-level qudit $q_1$ (including an auxiliary level $\ket{e}$ exploited during the gate) and a four-level qudit $q_2$, but the scheme can be straightforwardly extended to qudits of different size. As a unique knob to implement the gate, we exploit the tunability of the resonator frequency by insertion of proper SQUID elements. Such a tunability can reach about $30 \%$ of the bare frequency $\omega_0$  with a reduction of the resonator quality factor $Q$ of only a factor of 2-3 \cite{Palacios-Laloy2008} and can be as fast as 1 ns \cite{Wang2013}.
\\
We then proceed along the following steps (see Fig.~\ref{fig:scheme_2q}):
\begin{enumerate}[(a)]
    \item In the idle phase the resonator frequency is set 
    off-resonance from all qudit excitations (Fig. \ref{fig:scheme_2q}-a) and no photons are present (dark resonator).
    \item We then bring the resonator into resonance with $E_{0 1}^{(q_2)}$, i.e. the $\ket{0}\leftrightarrow\ket{1}$ transition of $q_2$ (Fig. \ref{fig:scheme_2q}-b). If state $\ket{1}$ of $q_2$ is populated, a photon is emitted and the resonator becomes bright (panel c). 
    \item We then induce a $2 \pi$ transition from $\ket{0}$ to $\ket{e}$ on $q_1$ (Fig. \ref{fig:scheme_2q}-c) with photon absorption and re-emission. If the transition is {\it resonant}, the amplitude is completely transferred from $\ket{0}$ to $\ket{e}$ and back, resulting in  
    a $\pi$ phase added to the $\ket{01}$ component of the two-qudit wave-function. If we slightly detune the excitation from resonance (the so-called {\it semi-resonance}), only part of the amplitude is temporarily excited to $\ket{e}$ and comes back with an additional phase which can be chosen by setting the detuning \cite{SciRep15}. In this way, we add an arbitrary phase  to $\ket{01}$.     
    \item Finally (Fig. \ref{fig:scheme_2q}-d), we bring again the resonator into resonance with $E_{0 1}^{(q_2)}$. 
    For the initial component $\ket{1}$ on $q_2$ (which had emitted the photon in panel b) the photon is absorbed again and we go back to $\ket{01}$ with no photons in the resonator and the additional phase inherited from the previous step. 
    The time interval between steps (b) and (d) can be adjusted in order to compensate any additional phase acquired, so that only (c) introduces a two-qudit phase. This happens only for the $\ket{01}$ component, while all the others are left unaffected by the sequence.
    Hence, the whole sequence implements $U_\varphi^{01}$, as desired.
\end{enumerate}

We conclude this section by noting that, although significantly slower, the dispersive approach has some significant advantages which must be kept in mind. In particular, gates are practically insensitive to photon loss, which instead becomes more relevant in the resonant regime.
Hence, a quantitative comparison between the two methods is needed, as reported in Sec. \ref{sec:simulations} below.

\subsection{Readout}
\begin{figure}[t!]
    \centering
    \includegraphics[width=0.45\textwidth]{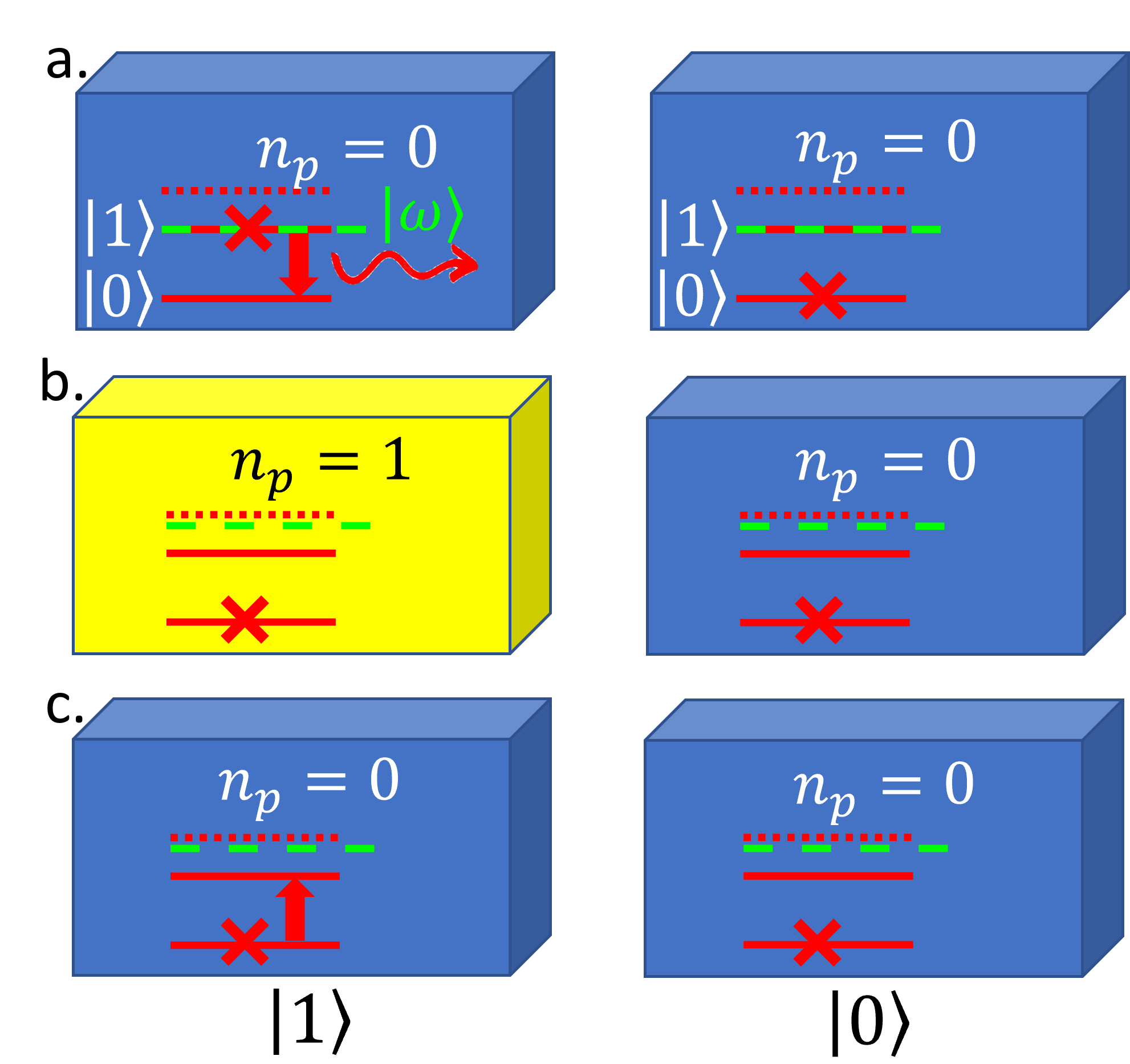}
    \caption{Scheme of the resonant readout of a qubit in state $\ket{0}$ (left) or $\ket{1}$ (right). a) Starting with no photons (dark cavity) the resonator is brought into resonance with $E_{01}$. If the qubit was in state $\ket{1}$ (left) a photon is emitted and the cavity becomes bright (b), bringing the qubit to the ground state. At this step, a single-photon counter detects the possible presence of the photon (and annihilates it). In case of positive outcome, a classical drive sends back the qubit state to $\ket{1}$ (c), thus implementing a projective measurement on the qubit. }
    \label{fig:scheme_ro}
\end{figure}

We provide here two different strategies to readout the final state of a qudit strongly coupled to a superconducting resonator. 
Analogously to two-qudit gates, these two strategies are designed to work either (i) in the dispersive or (ii) in the resonant regimes. \\
Case (i) (proposed in~\cite{GomezLeon2022}) mimics the approach followed with superconducting qubits and relies on transmission measurements in the dispersive regime. The resonator frequency experiences a shift $\chi$ that depends on the qudit state and that is proportional to the dispersive resonator-qudit coupling. More specifically, it scales as $(G_i^m)^2/\Delta$. 
Well-known advantages of this readout technique are its single-shot nature and that it is a quantum non-demolition measurement~\cite{Zueco2009}.  \\
The alternative approach (ii) we propose here is based again on resonant photon emission, as illustrated in Fig.~\ref{fig:scheme_ro}. 
In practice, it is equivalent to swap the quantum state we want to detect from the qudit to the resonator (by bringing a specific qudit gap into resonance with the photon field for a time $\pi/2G$), detect the possible presence of a photon in the resonator and swap the projected state back by a classical drive depending on the measurement outcome. This effectively implements a projective measurement on the qubit. Extension to the qudit case is straightforward, by addressing different spin gaps one at a time.

The readout fidelity of (i) depends on a proper discrimination of the resonator shift $\chi$. Hence, the main limitation comes from the broadening of the resonance peaks, mainly due to resonator losses and to spin-dephasing of the qudit, as discussed in the next subsection (\ref{subsec:deco}). This broadening must be smaller than the frequency separation associated with each qudit state. The effect can be partially mitigated by measuring the transmission phase rather than its modulus, as it seems to be more insensitive to broadening~\cite{GomezLeon2022}. \\
In addition, the time needed to complete the readout process scales as $1/\chi$. Readout times of the order of $50-100$ ns have been achieved with superconducting transmons having $\chi = 7.9$ MHz and operating at $G/\Delta \sim 1/10$, using high sensitivity detection stages, involving parametric amplification \cite{Walter2017}. Applying similar detection conditions, we expect $\chi$ to be approximately $2$ orders of magnitude smaller in our MSQP, with corresponding readout times of $\approx 5-10 \, \mu$s. For algorithms that do not involve 
any measurement dependent feedback during the execution 
process, the readout time needs to be sufficiently short as compared to the spin relaxation time $T_{1}$. This requirement does not pose a very stringent limitation, as $T_{1}$ becomes rather long at very low temperatures. Yet, if one attempts to implement quantum error correction codes, then the 
limiting time scale is the spin coherence time $T_{2}$, thus it 
might be necessary to look for faster methods. \\
The resonant approach (ii) provides such a large speedup. The latter will also increase the readout fidelity, by reducing the harmful effects of both spin decoherence and resonator losses. Indeed, the time required to implement a resonant readout is dominated by the time for photon emission ($\approx 250$ ns for the lowest energy gap of the $S=10$ molecule considered here), while single-photon detectors can work as fast as $50-100$ ns \cite{Wang2020}. \\

In summary, resonant and dispersive readout provide different ways to measure the state of the qudits. Each regime of operation comes with certain advantages and disadvantages, which is why different experimental setups might embrace different options.
The dispersive regime is non-demolition, but it is much slower due to the small dispersive spin-resonator coupling. In addition, it requires the design of more complex protocols, specially in the case of qudits with a large number of levels \cite{GomezLeon2022}. 
In terms of decoherence, it is mainly limited by spin dephasing. In contrast, the resonant readout is a projective measurement that can be applied to any given pair of levels. In this regime, the effect of photon loss is expected to increase, but the significant speedup which can be achieved makes it practically negligible. \\
In both cases, the most direct way to improve the readout fidelity is to increase the spin-photon coupling $G_i^m$, such that in scheme (i) transmission peaks are more separated and the readout time is shorter, whereas in scheme (ii) resonant photon emission becomes faster. 

\subsection{Decoherence}
\label{subsec:deco}
The most important decoherence channels in our platform are represented by (i) photon loss and single-molecule (ii) pure-dephasing, originating at low temperature from the interaction between the central spin and the surrounding nuclear spin bath \cite{npjQI}. \\
As already noted above, photon loss (occurring at a rate $\omega_r/2\pi Q$, with $Q$ the resonator quality factor) is a much more important effect in the resonant compared to the dispersive regime, due to real (rather than virtual) photon exchange. Nevertheless, by encoding the logical state within the $n_{p}=0$ subspace this effect is largely reduced and limited to the resonant emission/absorption processes of two-qudit gates and readout. In all other steps of computation (namely single qudit rotations and idle phases) photon loss is irrelevant.\\
As far as spin dephasing is concerned, a high-fidelity two qudit gate requires coherence time $T_2$ significantly longer than the gating time $\tau_{2q}$. With the present parameters $\tau_{2q} \sim \, \mu$s in the resonant regime, thus requiring $T_2>50~\mu$s to achieve high-fidelity gates.
Remarkably, these values have already been experimentally accomplished by proper chemical optimisation of the molecular structure, combined with deuteration and dilution, in V/Cu complexes \cite{Bader,Atzori_JACS}, reaching even the ms range \cite{Zadrozny}.
Moreover, it is important to note that the setup we are considering here will work (i) at much lower temperatures than those usually employed in $T_2$ measurements and (ii) at the single-molecule level, where a possible distribution of anisotropy orientations observed in a crystal will no longer be an issue. Hence, by working on properly engineered single molecules and in the mK region, coherence times approaching ms can be expected.  \\

\subsection{Scalability}
\label{subsec:scala}
Having introduced all the working principles of the elementary unit of the MSQP, we can now address the final scalability issue.  \\
One of the appealing features of this approach is that the MSQP can be scaled up at different levels. 
First, by chemical design that brings in multiple states within each molecular unit. Eventually, this would allow proof of concept implementations of qudit based algorithms even working with molecular crystals. Yet, moving beyond the Hilbert space provided by each molecule relies on the coupling to a superconducting circuit. Several molecular spin qudits can be integrated within a superconducting resonator, up to the limit allowed by nanofabrication techniques (as molecules themselves are microscopic entities). Finally, one can consider an array of capacitively-coupled resonators, each one containing a single or a few molecular qudits strongly coupled to the local photon mode, described by the boson field $\hbar \omega_{r,i} (a^\dagger_i a_i+1/2)$. The capacitive coupling between neighboring resonators is described (in the rotating wave approximation) by an interaction term of the form $\kappa a_i^\dagger a_j + {\rm h.c.}$, with $\kappa/2\pi$ in the $10-25$ MHz range \cite{Underwood2012}.
Neighbouring resonators are characterised by different bare frequencies (i.e. $|\omega_{0i}-\omega_{0j}|>> \kappa$), so that the inter-cavity photon hopping is effectively turned off in the idle phase. \\
To implement two-qudit gates between molecules placed into different resonators, 
we proceed in a way analogous to that illustrated in Fig. \ref{fig:scheme_2q}, but with an additional step. The two qudits we consider can now be located into neighboring resonators $i$ and $j$. Hence, after photon emission in step b), one needs to bring resonators $i$ and $j$ into mutual resonance, i.e. make $\omega_{0i}=\omega_{0j}$, thus inducing hopping of the single photon from the resonator containing qudit 2 to that containing qudit 1. We then proceed with photon absorption and re-emission as in Fig. \ref{fig:scheme_2q}-c) and repeat the previous steps of photon hopping and re-absorption, as in Fig. \ref{fig:scheme_2q}-d), eventually implementing the same $U_\varphi^{pq}$ gate illustrated above.\\

\section{System layout and technical aspects}
\label{sec:layout}
We now present in more detail the layout of the MSQP elementary unit, together with numerical estimates of the spin-photon coupling which can be accomplished by a proper design. 

\begin{figure}
    \centering
    \includegraphics[width=0.5\textwidth]{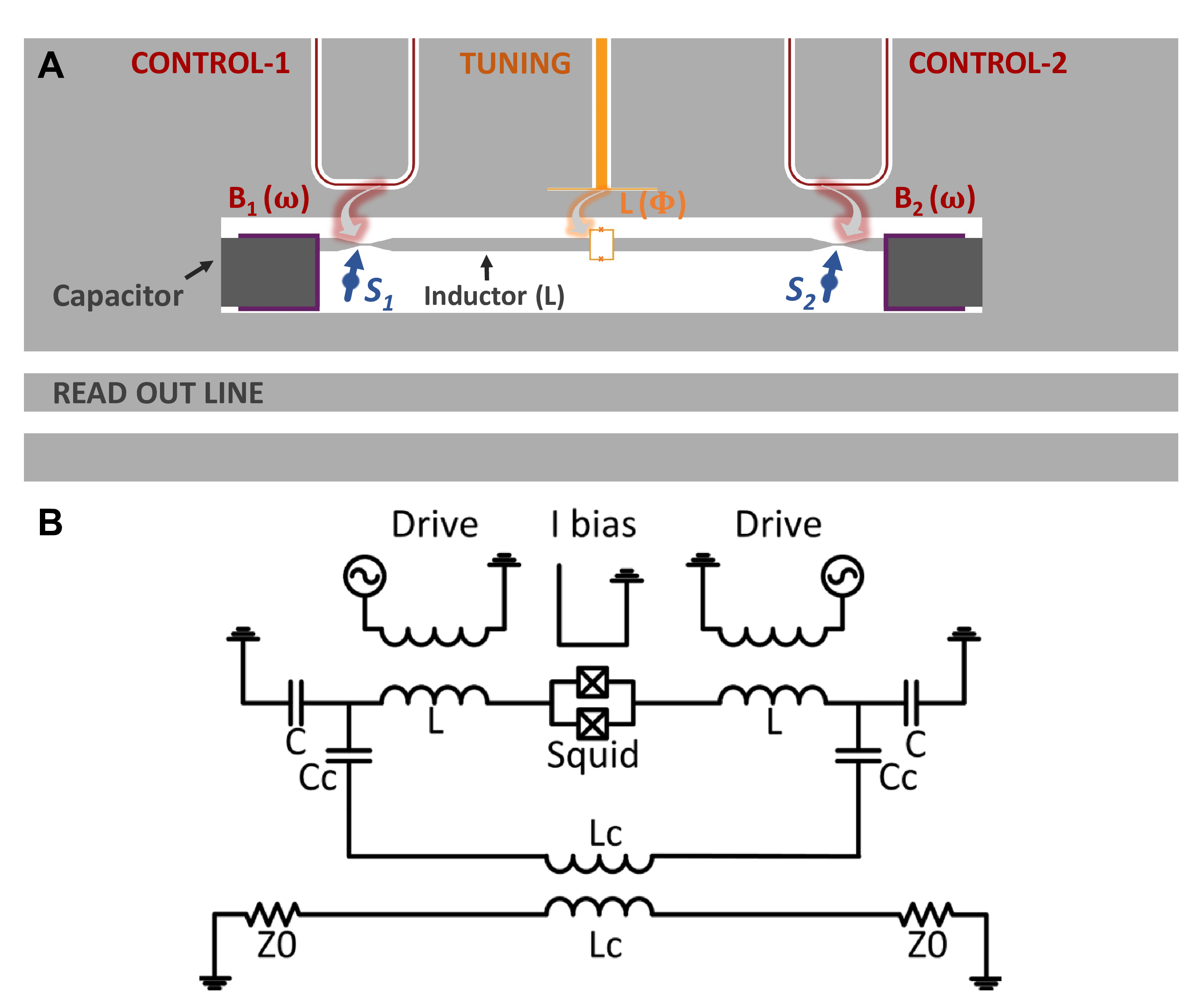}
    \caption{Schematic layout (A) and electric circuit (B) of the MSQP elementary unit, which consists of an on-chip $LC$ superconducting resonator with a few magnetic molecules deposited on top. Nano-sized constrictions concentrate the magnetic field at specific regions, where the coupling to single molecules, with spins ${\bf S}_1$ and ${\bf S}_2$ is enhanced. Transmission lines introduce electromagnetic pulses to coherently control each spin qudit and to tune the flux though a dc SQUID in series with the inductor. The latter serves to tune the resonator characteristic frequency. An additional transmission line reads out the state of the resonator, which provides information on the spin qudit states.}
    \label{fig:layout}
\end{figure}

The basic unit of our platform is schematically shown in Fig. \ref{fig:layout}. It consists of two molecular spin qudits coupled to a resonator, whose resonant frequency $\omega_{\rm r}$ can be tuned by means of a SQUID. As explained above, the resonator reads out the spin states and mediates an effective interaction between the two qudits that allows performing conditional quantum operations. The circuit includes also control transmission lines, which generate local magnetic field pulses to control the spin states of each qudit and the flux through the SQUID loop, and a readout line. Since the aim of this work is to discuss its operation and potential based on a realistic layout, we provide here estimations for all relevant parameters based on circuits that are feasible. 

As the previous sections show, the main technical challenge behind this proposal is the achievement of a  spin-photon coupling that is sufficiently strong as compared to the decoherence rates of the spins (typically determined by $1/T_2$) and the photons ($\omega_{\rm r}/2 \pi Q$). Values of $Q$ in the range $5\times 10^{5}-10^{6}$, corresponding to photon 
line widths of the order of $1-10$ kHz, can be achieved for $3-8$ GHz resonators \cite{Megrant2012}.\\

\begin{figure}
    \centering
    \includegraphics[width=0.45\textwidth]{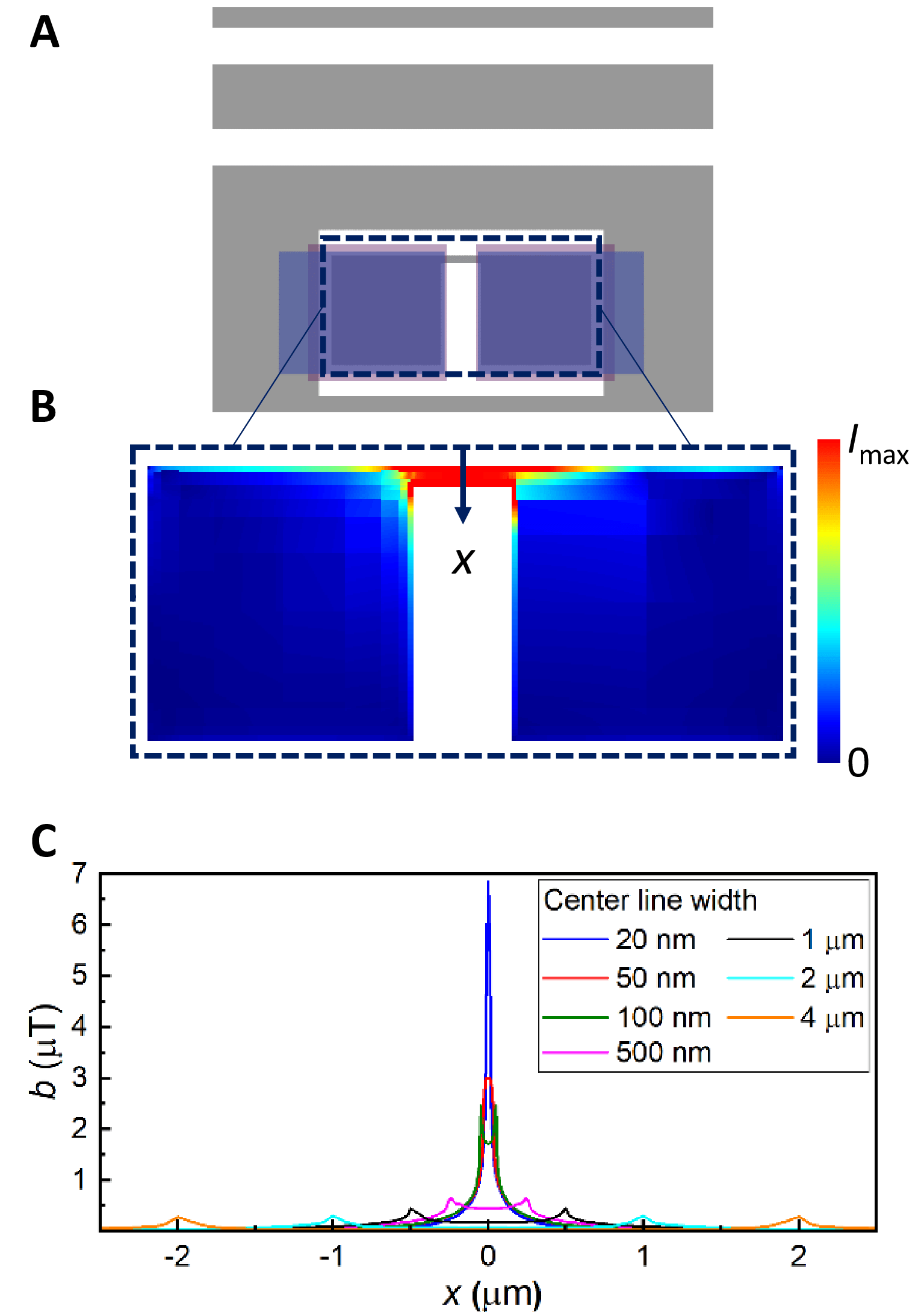}
    \caption{A) Layout of a $7.5$ GHz $LC$ superconducting resonator with a $22 \mu$m long and $4 \mu$m wide inductor and two parallel plate capacitors. B) Two dimensional plot of the distribution of currents at resonance. C) Spatial dependence of the microwave magnetic field generated by the resonator for different inductor widths $w$.}
    \label{fig:resonator}
\end{figure}

For this reason, we consider here a circuit designed to maximize and then also confine the superconducting current in resonance, thus enhancing the photon magnetic field in a nanoscopic region. 
The circuit consists of a lumped element $LC$ resonator, made of two capacitors connected by a single wire inductor (Fig. \ref{fig:resonator}A). These resonators provide ample room for tuning relevant parameters, such as impedance $L$ and resonance frequency, without compromising the transmission through the readout line that carries information about the system state. Moreover, several of them can be coupled to a single transmission line. 
Parallel plate capacitors are employed as they are free from parasitic inductances present in other designs such as interdigitated capacitors. This reduces the mode volume, confining the current $I$  almost completely in the inductor, as shown in Fig. \ref{fig:resonator}B. Near resonance, the photon energy splits equally between the magnetic and electric components. Since the former equals $LI^{2}/2$, a very low inductance leads to large current intensities for single photon excitations. The circuit shown in Fig. \ref{fig:resonator} has a $22 \, \mu$m long, $4 \, \mu$m wide and $100$ nm thick Nb inductor. We have simulated its electromagnetic response using the SONNET package \cite{SONNET}. This gives a resonator inductance $L\simeq 12.9$ pH, leading to a rms current $I_{\rm rms} = \sqrt{\hbar \omega_{\rm r}/2L} \simeq 438$ nA. In the simulation we have considered a kinetic inductance of $0.1$ pH/sq, corresponding to a $100$ nm Nb film. 

The photon magnetic field ${\bf b}$, which determines the coupling to the spins, is then calculated with the 3D-MLSI simulation package \cite{Khapaev2002}. The input parameters are the rms current ($438$ nA) flowing through the inductor and the London penetration depth of Nb $\lambda_{\rm L} \simeq 90$ nm. An example of the magnetic field intensity obtained at a distance of $1$ nm above the chip's surface is shown in Fig. \ref{fig:resonator}C. For the original inductor width $w = 4 \, \mu$m, it peaks at values below $0.4 \, \mu$T near the inductor edges.    

A way to further enhance $b$ at specific locations is to locally confine the current by fabricating nanoconstrictions in the inductor line  \cite{Jenkins2013}. Such constrictions can be made by ion-beam lithography and they do not significantly affect the resonator's properties provided that they are sufficiently short (below $1 \, \mu$m) \cite{Jenkins2013}. Figure \ref{fig:resonator} shows how $b$ increases as the constriction width $w$ is decreased, reaching values close to $7 \, \mu$T for $w=20$ nm, close to the limit of nanofabrication techniques.  

\begin{figure}
    \centering
    \includegraphics[width=0.45\textwidth]{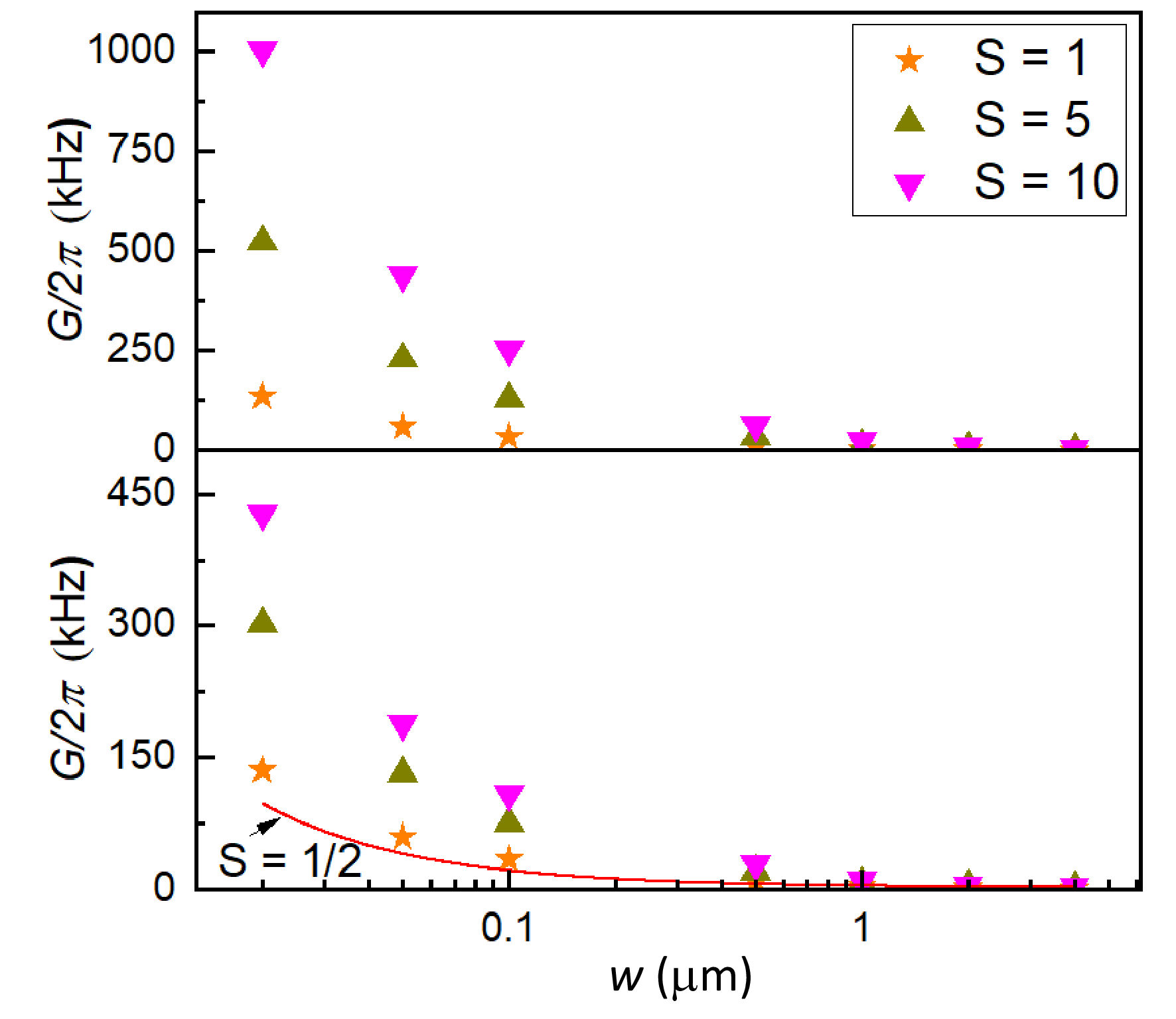}
    \caption{Coupling to photons of individual spin qudits, located $1$ nm above the the inductor line of the $LC$ resonator shown in Fig. \ref{fig:resonator}A. Results calculated for spins with easy axis ($D < 0$ in Eq. (\ref{eq:HamS}) and easy plane ($D > 0$) are shown in, respectively, the bottom and top panels. They correspond to a different spin ground state and, therefore, to different resonant spin transitions. The solid line gives the coupling obtained for a simple $S=1/2$ spin qubit.}
    \label{fig:coupling}
\end{figure}

Once the photon magnetic field is known, the spin-photon coupling $G^{m}$ in Eq. (\ref{eq:HamSp}) can be calculated as follows
\begin{equation}
G^{m} = \frac{g \mu_{\rm B}}{\hbar} \vert \langle m \vert {\bf b} \cdot {\bf S} \vert m+1 \rangle \vert
\label{eq:G1}
\end{equation}
The direct relationship between the $G^{m}$ and $b$ suggests that, near a nanoconstriction, the spin-photon coupling can 
reach values much higher than those achieved for conventional resonators. Following this route, spin-photon couplings $G/2\pi$ near $1$ kHz have been achieved experimentally for single spins $1/2$ located close to a $45$ nm constriction in a $1.5$ GHz co-planar resonator \cite{Gimeno2020}, whereas $G/2 \pi \simeq 3$ kHz was achieved with $8$ GHz lumped element resonators having $100$ nm wide inductors \cite{Ranjan2020}. Figure \ref{fig:resonator}c shows that working with very low impedance resonators and sufficiently high frequencies (in this example, $\omega_{\rm r}/2 \pi = 7.5$ GHz) much higher values can be envisioned. The simulations give $G /2 \pi \simeq 95$ kHz for a $S=1/2$ qubit located near a $20$ nm constriction in the inductor of the circuit shown in Fig. \ref{fig:resonator}.  
Besides, and as discussed above, the coupling is also enhanced when working with high spin molecules. The bottom and top panels in Fig. \ref{fig:coupling} correspond to spin qudits with easy-axis ($D<0$ in Eq. (\ref{eq:HamS})) and easy-plane ($D>0$) magnetic anisotropy, respectively. It follows that the latter case, characterized by a ground state with $m=0$, represents a more favourable situation. Then, single spin to single photon couplings as large as $1$ MHz can be achieved with a $S=10$ molecule. A further enhancement could be obtained by proper choice of the molecule (e.g., with larger $S$) or by choosing peculiar eigenstates, such as
 atomic-clock transition states $(\ket{S} \pm \ket{-S})/\sqrt{2}$ \cite{Jenkins2013,Jenkins2016}. Nonetheless, 
this value is already close to the physical limit for this approach, as it involves an optimum circuit design combined with a close to maximum photon confinement and a close to perfect integration of the molecules with the device.  

The latter condition might at first look daunting. Molecules can be deposited from solution onto the circuits and then delivered to the region of the nanoconstrictions by an atomic force microscope \cite{Gimeno2020} or by exploiting the tendency of some molecules to self-organize on surfaces \cite{Domingo2012,Serrano2020,Gabarro2022,Tesi2023}. Yet, depositing single molecules sufficiently close to each constriction is very challenging and one will often end up having several molecules not too far apart. Nevertheless, we can address and manipulate individually each molecule (or each pair of molecules for two-qudit gates) by exploiting (i) the magnetic anisotropy (Eq. (\ref{eq:HamS}))  and (ii) the strong dependence of the spin-photon coupling on the precise location (see Fig. \ref{fig:resonator}C). Indeed, in general we expect different molecules to deposit with a different orientation. Hence, only a specific one will meet the resonance condition required to implement both single- and two-qudit gates (as explained in Sec. \ref{sec:wp}), while off-resonant ones will not undergo any evolution. Furthermore, a reduction of the coupling of at least one order of magnitude for molecules placed $50$ nm apart from the nanoconstriction will practically forbid two-qudit gates (section IIC) and make these spins undetectable (section IID). 
Therefore, one can take advantage of the statistical nature of the molecular deposition processes and select those molecules from the ensemble that fulfill a given condition vis a vis a given circuit design. The molecular integration requisite then softens to having a sufficiently sparse surface coverage, which is achievable with molecular evaporation or self-organization techniques \cite{Serrano2020,Tesi2023}. 

\section{Numerical Simulations}
\label{sec:simulations}
Having introduced both the working principles and the realistic set of experimental conditions to operate the  MSQP, we can now assess and quantify its performance by numerical simulations. 
To this aim, we consider the full series of operations (i.e. classical pulses and variations of the resonator frequency) to implement sequences of one- and two-qudit gates in our platform. 
As in the previous discussion, we focus for simplicity on a elementary unit of the scalable setup, i.e. a single resonator containing a pair of qudits.
Simulations are realised by numerically integrating the Lindblad equation for the full system density matrix $\rho$:
\begin{eqnarray} \label{eq:Lindblad}
    \dot{\rho} &=& -i[H+H_1(t), \rho]  \\ \nonumber
    &+& \frac{1}{T_2} \sum_{i} \left( 2 S_{zi} \rho S_{zi} - S_{zi}^2 \rho - \rho S_{zi}^2 \right) \\ \nonumber
    &+& \frac{\omega_r}{2 \pi Q} \left( 2 a \rho a^\dagger - a^\dagger a \rho - \rho a^\dagger a  \right),
\end{eqnarray}
where the first line describes the coherent evolution, the second models pure dephasing (where we have assumed the same $T_2$ for the two qudits) and the last one represents photon loss.
We assume in the following $G_i/2 \pi\approx 90$ kHz (slightly different for the two spins to better match experimental conditions), in line with the best values reported in Fig. \ref{fig:resonator}, and $\omega_0/2\pi=7.5$ GHz. 
This choice yields $G_i^0/2 \pi \approx 1$ MHz for transitions involved in two-qubit gates reported below. As far as the spin system is concerned, we use $D_1/2 \pi=7.1$ GHz, $D_2/2 \pi=7.7$ GHz, $g_i=2$, $B=50$ mT. The small static field is compatible with superconducting circuitry technology.
\begin{figure*}[t!]
    \centering
    \includegraphics[width=\textwidth]{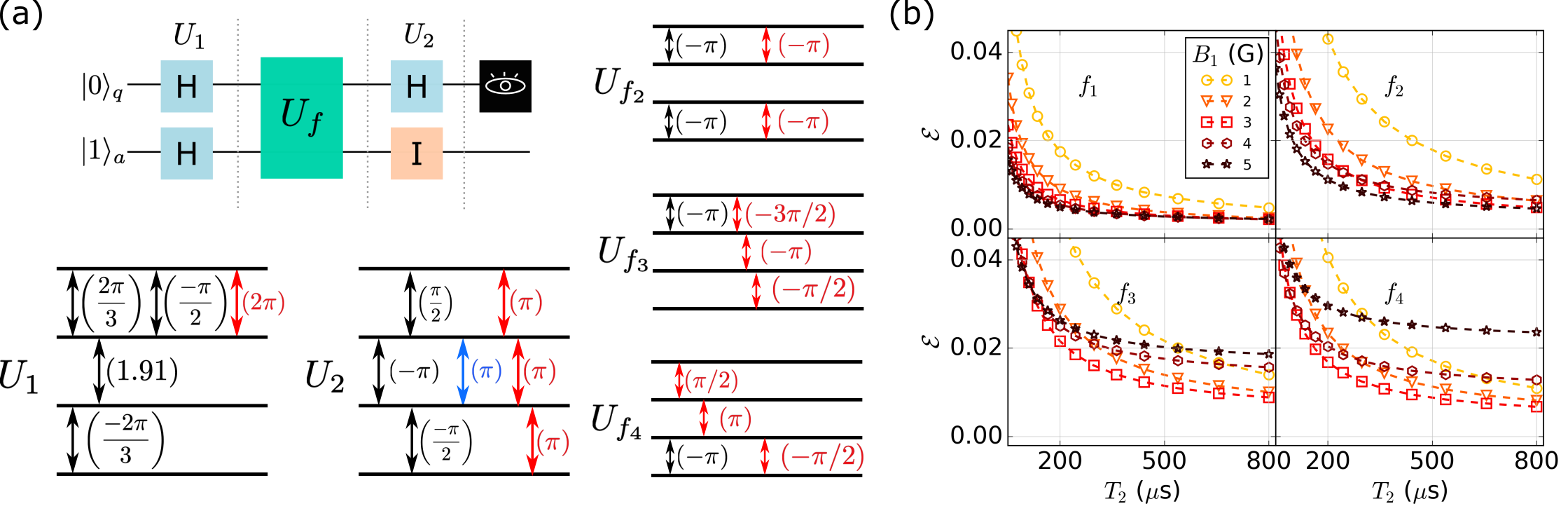}
    \caption{Decomposition of the circuit for the Deutsch-Jozsa (2 qubit) algorithm into a sequence of pulses on a 4-level qudit. (a) Schematic representation of the 2-qubit circuit and of the sequence of qudit operations needed to implement each unitary. Each arrow colour represents one rotation direction (parameter $\phi$ in Eq. \eqref{eq:IntUnitary}): black for Y, blue for X and red for Z. (b) Error dependence with the driving magnetic field, $B_1$, and with the decoherence time $T_2$. The error is computed as $\mathcal{E} = Tr\left(\rho P\right)$ and $P$ is the projector onto the subspace in which the final wave function must not be contained according to the applied function, $U_{f_i}$.}
    \label{fig:DJsketch}
\end{figure*}
Below we first report simulations of single qudit operations, using classical driving fields to implement the Deutsch algorithm on a 4-level qudit as illustrated in Sec. \ref{subsec:1qudit}. We then quantify and compare the performance of two-qubits entangling gates such as the controlled-Z and $i$SWAP gates described above, employing the resonant and dispersive approaches, respectively. We finally focus on specific applications such as quantum simulation, in which several one- and two-qubit gates are concatenated.

\subsection{Multi-qubit algorithms onto a single qudit}
Hereafter we exploit the four lowest energy states of qudit 1 to encode two qubits and we show how to decompose the two-qubit Deutsch-Jozsa (DJ) algorithm \cite{DeutschOriginal} into a sequence of planar rotations on the qudit. The aim of this algorithm is to determine if a given function, $f$, is either constant or balanced. The algorithm implements the transformation $|x\rangle_q|y\rangle_a \rightarrow |x\rangle_q|y \oplus f(x)\rangle_a$ producing $|0\rangle_q$ as the final state for the input qubit if $f$ is constant and $|1\rangle_q$ otherwise. Here $|\ \rangle_{q(a)}$ denotes the state of the input (ancilla) qubit and the sum is a mod(2) addition. The constant functions that we are going to implement are $U_{f_1} = \mathbb{I}^{\otimes2}$ and $U_{f_2} = \mathbb{I}_q\otimes X_a$ and $U_{f_3} = CX_{q\rightarrow a}$, $U_{f_4} = X_q CX_{q\rightarrow a} X_q$ are the balanced ones. Here $CX_{q\rightarrow a}$ is a controlled-NOT gate where $q$ acts as a control on $a$. Following the procedure detailed in Sec. \ref{subsec:1qudit}, we map the following levels from our qudit $\{|m\rangle\} = \{|0\rangle, |+ 1\rangle, |-1\rangle, |-2\rangle \} \equiv \{|0\rangle, |1\rangle, |2\rangle, |3\rangle \}$ to the multi-qubit ones: $\{|0\rangle, |1\rangle, |2\rangle, |3\rangle \} \equiv \{|0\rangle_q|0\rangle_a, |0\rangle_q|1\rangle_a, |1\rangle_q|0\rangle_a, |1\rangle_q|1\rangle_a \}$. 
With this relation, we can now proceed to decompose the multi-qubit circuit into the exact sequence of pulses that have to be applied to our molecule to implement the full algorithm, which is shown schematically in Fig. \ref{fig:DJsketch} (a). It is important to notice that the levels chosen from our molecule do not have the ideal ladder connectivity since the states $|1\rangle$ and $|2\rangle$ are not connected directly. This is solved by adding a $|0\rangle\rightarrow|1\rangle$ gate to the $|0\rangle\rightarrow|2\rangle$ rotation.
The operations that appear as arrows in Fig. \ref{fig:DJsketch} (a) represent the different rotations to be applied: the colour indicates the axis direction and the parameter in brackets the rotation angle (parameters $\phi$ and $\theta$ in Eq. \eqref{eq:IntUnitary}, respectively). Recall that rotations around the $z$ axis can be implemented from the other two axis rotations [Cf. Sec. \ref{subsec:1qudit}]. With this we now evolve the system integrating Eq. \eqref{eq:Lindblad} for different values of the driving magnetic field, $B_1$, and of the decoherence time, $T_2$, to compute the error committed in the implementation. In the qubit case one expects to get the state $|0\rangle_q$ ($|1\rangle_q$) for a constant (balanced) function. In the qudit case, with this encoding, measuring $|0\rangle_q$ ($|1\rangle_q$) corresponds to a projection onto the subspace spanned by the states $\{|0\rangle, |1\rangle\}$ ($\{|2\rangle, |3\rangle\}$). Therefore, to evaluate the performance of our implementation, we compute the error as $\mathcal{E} =  Tr(\rho P)$, being $\rho$ the final state obtained from the pulse sequence and $P$ the projector corresponding onto the "wrong" subspace, i.e. $P = |0\rangle\langle0| + |1\rangle\langle1|$ for $U_{f_3}$ and $U_{f_4}$; $P = |2\rangle\langle2| + |3\rangle\langle3|$ for $U_{f_1}$ and $U_{f_2}$. Fig. \ref{fig:DJsketch} (b) shows that the error is clearly reduced by increasing $T_2$ and reaches values of the order of 1 \%. Increasing $B_1$ has the effect of speeding-up the gates, thus reducing the effect of dephasing. However, increasing it too much can worsen the results due to leakage effects.

\begin{figure}[b!]
    \centering
    \includegraphics[width=0.5\textwidth]{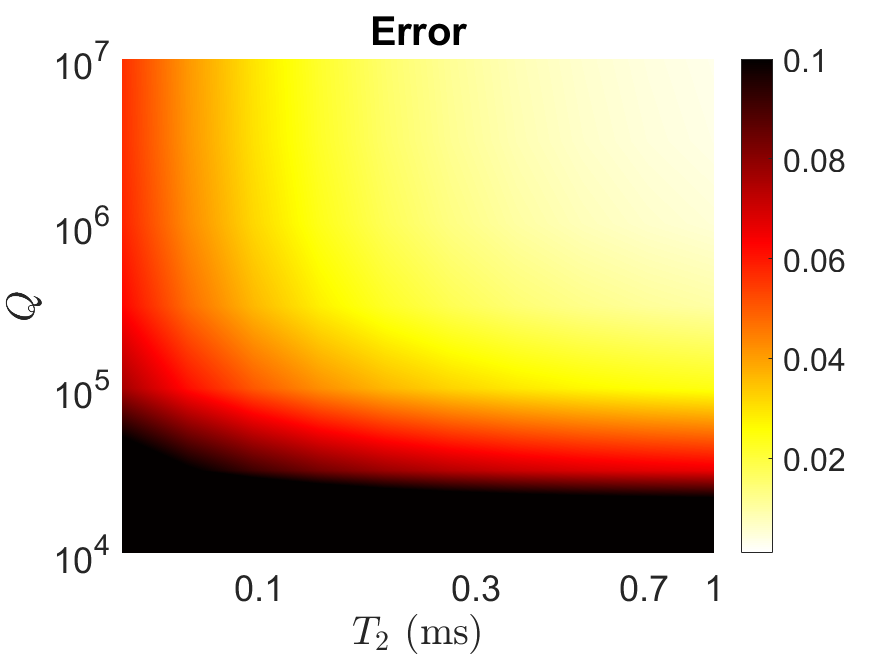}
    \caption{Error $1-\mathcal{F}$ (log-scale) in the implementation of the resonant controlled-Z gate as a function of the resonator quality factor $Q$ and of the qudits coherence time $T_2$. Random components on $\ket{p}\ket{q}$ initial state are given by $\ket{0}\ket{q}=(0.31,0.46,0.48,0.37)$ and $\ket{1} \ket{q}=(0.37,0.25,0.25,0.24)$ with $q = 0,1,2,3$.}
    \label{fig:ErrorCZ}
\end{figure}
\subsection{Two-qudit gates}
\begin{figure}[t!]
    \centering
    \includegraphics[width=0.48\textwidth]{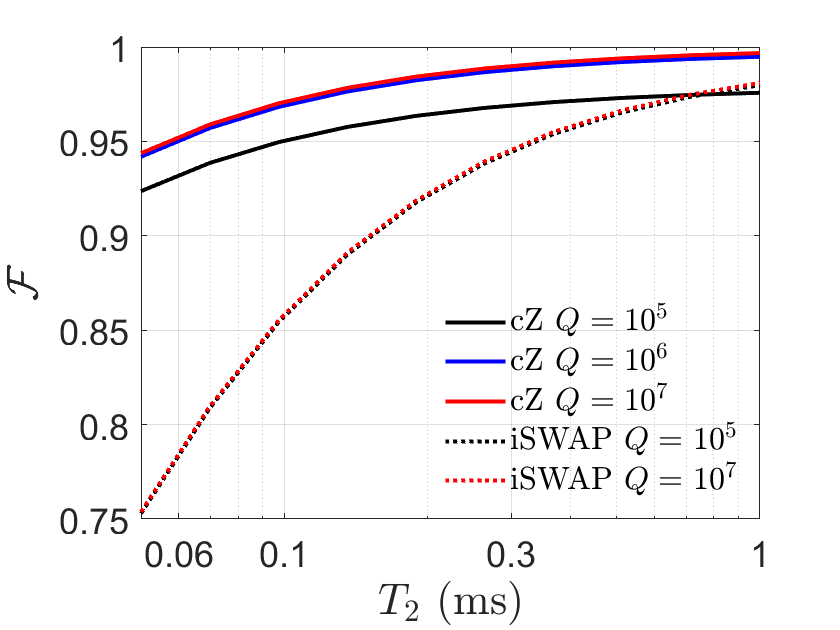}
    \caption{Fidelity in the implementation of the controlled-Z (cZ) gate in the resonant regime (solid lines) and of the $i$SWAP gate in the dispersive regime (dotted) as a function of the qudits coherence time $T_2$, for different values (colors) of the resonator quality factor $Q$. }
    \label{fig:FidelityCZ_T2}
\end{figure}

We now investigate the implementation of a two-qudit controlled-Z gate according to our novel
resonant approach, between a two- and a four-level qudit (plus an auxiliary level $\ket{e}$ used during the gate), as in the scheme of Fig. \ref{fig:scheme_2q}.
As figures of merit for the performance of the gate, we consider here the fidelity $\mathcal{F}=\braket{\psi | \rho | \psi}$, which quantifies the overlap between the desired state $\ket{\psi}$ that one gets in an ideal implementation of the gate and the actual state $\rho$ obtained by solving Eq. \eqref{eq:Lindblad}. 
As a benchmark, we consider a particularly error-prone initial state, in which all the components in the computational basis are populated (randomly) with significant weight (see caption of Fig. \ref{fig:ErrorCZ}).  \\
Fig. \ref{fig:ErrorCZ} shows a colormap of the resulting error $\mathcal{E}=1-\mathcal{F}$ as a function of the resonator quality factor $Q$ and of the qudits coherence time $T_2$.
The error is clearly reduced both by increasing $Q$ and $T_2$, but it shows a stronger dependence on the qudit coherence time. In particular, $Q$ values above $10^6$ give only a minor improvement, while $\mathcal{F}$ smoothly increases with $T_2$ even in the ms range. Nevertheless, the fidelity already overcomes $99 \%$ for $T_2 = 400 \, \mu$s. This is made even more evident by considering the $T_2$ dependence of the fidelity, shown in Fig. \ref{fig:FidelityCZ_T2} for different values of $Q$. \\
As a comparison, here we also report simulations of the $i$SWAP gate in the dispersive regime, where we have assumed $\Delta_1/2 \pi = 20$ MHz, $\Delta_2/2 \pi = 30$ MHz for the lowest-energy transitions of the two spins in the idle configuration. These are then  tuned into mutual resonance ($\Delta_1 = \Delta_2 = 20$ MHz $ \times 2 \pi$) to activate the $i$SWAP. In this configuration, $\Delta_1 \approx 20 \, G_i^0$ and hence we are safely in the dispersive regime, with negligible leakage to the excited one-photon states. \\
We first note that the fidelity of the $i$SWAP gate is practically independent of $Q$, in contrast to the resonant case. This was expected, because of the virtual vs real exchange of photons in the two methods. However, the approximately 6 times larger value of $\tau_{2q}$ for implementing the $i$SWAP compared to the controlled-Z makes the $i$SWAP more prone to decoherence, resulting in a significantly lower fidelity, only approaching that of the controlled-Z at very long $T_2$. 

\subsection{Application: quantum simulation}
To better highlight the difference between the resonant and dispersive approach in implementing a conditional two-qudit dynamics, we consider here a specific application of the gates described above.
Indeed, different gates can be targeted to different goals. Therefore, depending on the specific algorithm, a more efficient decomposition can be found by exploiting either $i$SWAP or controlled-Z gates. 
A clear comparison can be made by considering the quantum simulation of the Heisenberg model, which can be decomposed by exploiting an equal number of controlled-phase or $U_{XY}(\tau)$ gates. 
In particular, our aim is to simulate the dynamics associated to the target Hamiltonian 
\begin{equation}
    \mathcal{H}_H = J \left( s_{x,1} s_{x,2} + s_{y,1} s_{y,2} + s_{z,1} s_{z,2}. \right) 
    \label{eq:Heis}
\end{equation}
where $s_{\alpha,i}$ are spin 1/2 operators. Performing a quantum simulation corresponds to implement the unitary $U(Jt) = e^{-i \mathcal{H}_H t}$, starting from a generic state. 
In the present case, the three terms in Eq. \eqref{eq:Heis} commute, and hence we can reproduce the dynamics of $U(Jt)$ by subsequently implementing the three unitaries $U_{\alpha\alpha}(Jt) = {\rm exp} \left[  -i Jt s_{\alpha,1} s_{\alpha,2}  \right]$ for $\alpha = x,y,z$. These can be obtained by combining either the controlled-phase ($U_\varphi$) or the $U_{XY}(\tau)$ with proper single-qubit rotations. \\
As far as the resonant controlled-phase is concerned, we start from $U_{zz}(Jt)$. This can be easily re-written in terms of controlled-phase gates and single-qubit rotations \cite{SciRep15}, i.e.  $U_{zz}(\varphi) \propto e^{i\varphi/2} U_{\varphi}^{11} R_z(\varphi/2)$. Here  $R_z(\varphi)$ are single-qubit rotations about the $z$ axis of an angle $\varphi$ on both qubits. By setting $\varphi = Jt$ we simulate the target $U_{zz}(Jt)$. The latter can be easily transformed into $U_{xx}(Jt)$ by a simple change of reference frame, i.e. $U_{xx}(Jt) = R_y(\pi/2) U_{zz}(Jt) R_y(-\pi/2)$ and analogously for $U_{yy}(Jt)$. Overall, we need three controlled-phase gates and five rotations (on both qubits) to simulate the Heisenberg model. 
\begin{figure}[b!]
    \centering
    \includegraphics[width=0.48\textwidth]{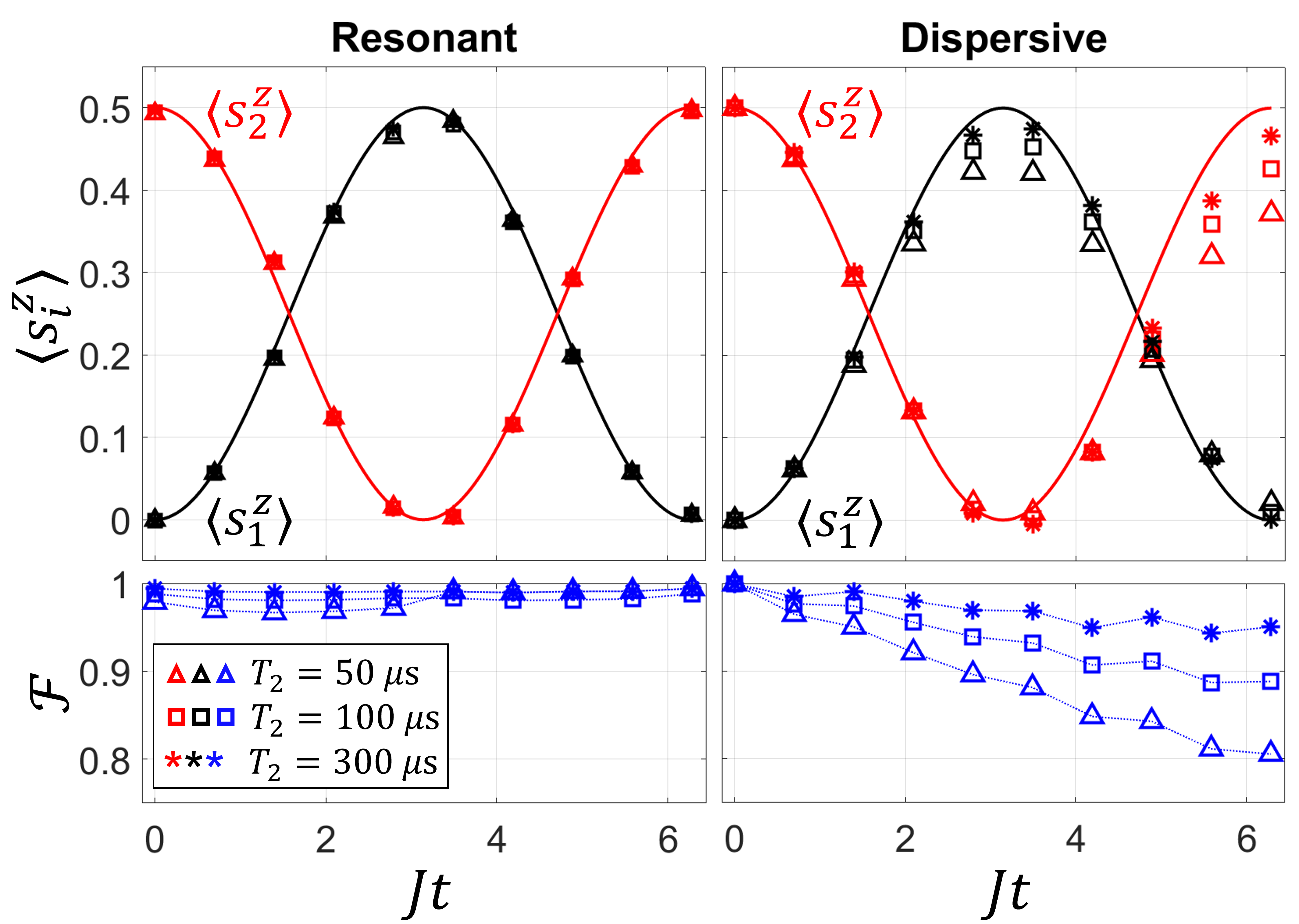}
    \caption{Comparison between resonant (a) and dispersive (b) approaches in the digital quantum simulation of the Heisenberg model, and associated fidelity (c,d) as a function of the simulated time. The system was initialized in $(\ket{00} + \ket {01})/\sqrt{2}$. In panels (a,b) we report the time evolution of the expectation values of $s_{z1}$ (red) and $s_{z2}$ (black). The resonator quality factor is fixed to $Q=10^6$.}
    \label{fig:qsHeis}
\end{figure}\\
Following instead the dispersive approach, $U_{XY} (\tau)$ is naturally mapped into $U_{xx}U_{yy}$. Then, we can exploit again 
changes of reference frame to transform this into $U_{xx}U_{zz}$ (by means of $R_x(\pm \pi/2)$ rotations) or into $U_{yy}U_{zz}$ (via $R_y(\pm \pi/2)$). By concatenating these three steps we get $U(Jt) = U_{zz}(Jt/2) U_{yy}(Jt/2)U_{xx}(Jt/2) U_{zz}(Jt/2)U_{xx}(Jt/2) U_{yy}(Jt/2) $ \cite{Solano2014,Chem}. Also in this case, the Heisenberg evolution is obtained by using three two-qubit gates $U_{XY} (\tau)$ and four single qubit rotations. Hence, 
this is a very good benchmark to compare the two methods. 
\\

To perform the quantum simulation, we exploit the two lowest energy levels of both qudits in the processor as logical states and the third one on $q_1$ as an auxiliary state for the implementation of the controlled-phase gate [see Sec. \ref{subsec:2qubits}].
Results of our simulations following the two different approaches are shown in Fig. \ref{fig:qsHeis}, where we plot the time evolution of the expectation values of $s_{z1}$ (red) and $s_{z2}$ (black).
First, it is worth noting that the resonant approach gives optimal results already for $T_2 = 50 \,\mu$s, in the whole range of simulation times $t$. Then, while for short $t$ the accuracy of the two methods (and the associated fidelity, lower panels) is similar, the performance of the dispersive approach breaks down as long as $Jt$ increases. Conversely, the resonant method displays a fidelity which is practically independent of $Jt$. This behavior can be easily understood by considering that $Jt$ is associated either to the angle $\varphi$ of the controlled-phase or to the time evolution $\tau$ of the $U_{XY}(\tau)$ gate. While the duration of the controlled-phase is only weakly dependent on  $\varphi$, the duration of $U_{XY}(\tau)$ increases linearly with $\tau$. As a consequence, in the resonant regime the simulation lasts approximately $3 \;\mu$s independently of the simulated time, while in the dispersive regime it increases up to  $17 \;\mu$s at $Jt=2\pi$, thus giving reliable results only for $T_2$ of hundreds of $\mu$s.

Having demonstrated the general better performance of the resonant approach, we now consider its application to another, more difficult problem represented by the quantum simulation of the transverse field Ising model (TIM). The target Hamiltonian on a chain of $N$ spins 1/2 is given by:
\begin{equation}
    \mathcal{H}_T = J \sum_{i=1}^{N-1} s_{z,i} s_{z,i+1} + b \sum_{i=1}^N s_{x,i}.
    \label{eq:TIM}
\end{equation}
The associated dynamics is nontrivial and gives rise to a quantum phase transition for specific values of the parameters $J$ and $b$. 
In the present case, the two non-commuting terms in Hamiltonian \eqref{eq:TIM}, namely $\mathcal{H}_1 = b \sum_{i=1}^N s_{x,i}$ and $\mathcal{H}_2 = J \sum_{i=1}^{N-1} s_{z,i} s_{z,i+1}$, require a Suzuki-Trotter decomposition to approximate the unitary $U$, i.e. 
\begin{equation}
    U = e^{-i\mathcal{H}_T t} \approx \left( e^{-i \mathcal{H}_1 t/n} e^{-i \mathcal{H}_2 t/n} \right)^n.
\end{equation}
Here $n$ is the number of Suzuki-Trotter steps in the decomposition, the first (one-body) term directly corresponds to single-qubit rotations $e^{-i \mathcal{H}_1 t/n} \equiv R_x(bt/n)$ on the whole register, while the second is obtained as described above.
\begin{figure}[t!]
    \centering
    \includegraphics[width=0.48\textwidth]{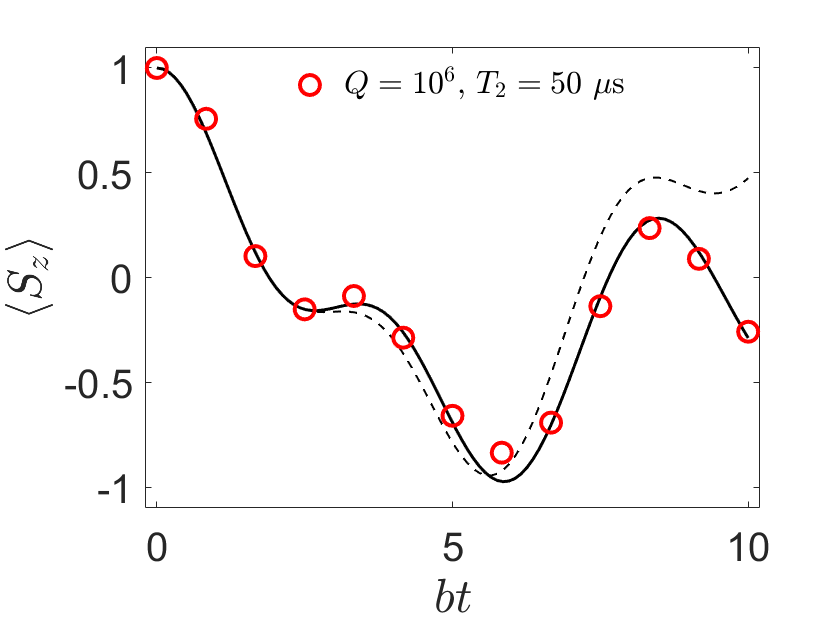}
    \caption{Quantum simulation of the time evolution of the system magnetisation $\langle S_z \rangle \equiv \langle s_{z,1}+s_{z,2} \rangle$ for the transverse field Ising model on two qubits. Dashed (continuous) line represent the exact evolution (after Trotter decomposition, $n=6$). Colored points are the results of full numerical simulation, including the effect of photon loss, parametrized by the resonator quality factor $Q$, and of spin pure dephasing, parametrized by the coherence time $T_2$. }
    \label{fig:QS_TIM}
\end{figure}
For our benchmark, we focus on a two-qubit system ($N=2$) in the most demanding regime $J=2b$, where the commutator between the two terms in the target Hamiltonian \eqref{eq:TIM} $[\mathcal{H}_1,\mathcal{H}_2]$  is the largest and hence a large number of Suzuki-Trotter steps is required for a reliable quantum simulation. \\
Results of our numerical simulations are reported in Fig. \ref{fig:QS_TIM}, using $n=6$ Trotter steps. This corresponds to a sequence of $n$ controlled phase gates (implemented by variations of the resonator frequency, as illustrated in Sec. II.C) and $2n$ rotations (implemented via classical drives). 
To compare the results of our simulation with the expected ones, we compute the final magnetisation of the target system $\langle S_z(t) \rangle = \langle \psi(t) | s_{z,1}+s_{z,2}| \psi(t) \rangle$, where $\ket{\psi(t)} = U \ket{\psi_0}$ is the target system state at the simulated time $t$ and $\ket{\psi_0} = \ket{00}$. \\
In spite of the significant number of operations (12 rotations and 6 controlled-phase gates) involved in the quantum simulation of each point in Fig. \ref{fig:QS_TIM}, the MSQP is able to reproduce very well the expected behavior of $\langle S_z \rangle$, even with $Q=10^6$ and $T_2 = 50 \, \mu$s. 

\section{Discussion and conclusions}
\label{sec:discussion}
We have introduced the working principles of a quantum processing unit based on molecular nanomagnets coupled to superconducting circuits, and tested its performance via numerical simulations based on a realistic experimental layout. The results are promising and indicate that the MSQP designed in this work traces a viable path for manipulating individual molecular qudits and for wiring them up in a scalable device. In particular, we exploit the hybrid character of the proposed architecture that combines the mobility of photons to the multi-level structure of spin systems. This allows one to implement entangling gates even between distant molecules in the register with no need of highly demanding SWAP operations, and to switch off completely their coupling at the end of the gate, thus avoiding crucial cross-talking issues which occur in presence of a permanent qubit-qubit interaction \cite{modules}.  

The simulations achieve a high fidelity in the implementation of two-qudit gates in the resonant regime,
by using spin coherence times of a few hundreds of $\mu$s. These results represent an important improvement compared to the dispersive regime, thanks to the significant speed-up achieved in the implementation of entangling gates. A drawback of the resonant approach (which is however strongly limited by encoding logical states in $n_{ph}=0$ subspace) is represented by photon loss, whose effect is instead practically negligible for the dispersive method. To mitigate these issues, intermediate regimes (with detuning of the order of $\sim 5 \, G_i^m$) could be explored. Although less flexible (the time evolution of $U_{XY}(\tau)$ should be limited to fixed values of $\tau$ to avoid leakage), this would allow for reaching gating times similar to the resonant case, with a reduced effect of photon loss. \\
Moreover, to solve specific problems (e.g. quantum simulation)  the MSQP could already operate with good performance even with significantly shorter values of $T_2 \sim$ tens of $\mu$s.
These values have already been achieved and even largely overcome in molecular spin qubits \cite{Bader,Zadrozny}, in much worse experimental conditions than those proposed here, i.e. at significantly higher temperatures and on a (diluted) molecular ensemble. The possibility to reach these $T_2$ with qudits needs still to be experimentally investigated, although theoretical results indicate that optimal coherence could be obtained in compounds with the suitable pattern of interactions \cite{npjQI,Chiesa2022}.

Further improvements could be obtained by enhancing the ratio $T_2/\tau_{2q}$ between coherence and gate times. This, in turn, calls for a joint multidisciplinary 
effort to increase not only spin coherence but also the coupling of spins to superconducting resonators. Increasing the values 
estimated under section \ref{sec:layout} by pushing the circuit miniaturization is a challenge, although the application of ultrahigh 
resolution nanolithography methods might still provide room for some improvement \cite{Lewis2019}. The alternative is to look for a 
different coupling regime between spins and photons. A possibility is to introduce interaction mediators based on magnonic excitations \cite{Fukami2021,Yuan2022}, e.g. in magnetic films or $2D$ materials,
investigate the coupling of spins with electric instead of magnetic fields (as recently done in an EPR setup \cite{George2015,Liu2019,Liu2021}) or even explore hybrid circuits that might bring tools from molecular electronics \cite{Cleuziou2006,Urdampilleta2011,Godfrin2017} into the circuit-QED realm. 
In parallel to seeking new circuit designs, the spin-photon coupling could also be improved via a proper engineering of the system Hamiltonian. For instance, a further factor of $\sim 2$ in the effective coupling could be reached by encoding qubits 
into atomic-clock transition states  $(\ket{S} \pm \ket{-S})/\sqrt{2}$, coupled by $S_z$ and protected from magnetic field fluctuations. \\ 
On the chemical side, different classes of compounds could be explored. Besides multi-spin clusters with very large total spin ground state \cite{Waldmann2008}, other possible implementations are offered by lanthanide-qubits, whose $g$ tensors can be engineered to achieve values up to 5-10 \cite{ErCeEr} and hence a corresponding enhancement of the spin-photon coupling. For instance, Yb(trensal) complex \cite{jacsYb} shows $g_z = 4.3$ and ensembles of this molecule have already been coupled to superconducting resonators \cite{Rollano2022}. \\
All these classes of compounds share the typical trademark of MNMs, i.e. they provide many low-energy states which represent a crucial resource to simplify quantum algorithms and especially to embed quantum error correction. By combining this peculiar advantage with an accurate design of the device, we have indicated here a clear route for the actual realization of a new promising kind of quantum chip. Indeed, integration of  molecular spins, offering novel advantages as multi-level logical units, into existing superconducting resonators takes advantages from both classes of materials and bridges the gap between current and future technologies.

\acknowledgements
This work has received funding from the European Unions Horizon 2020 research and innovation programme (FET-OPEN project FATMOLS) under grant agreement no. 862893.
Project also funded under the National Recovery and Resilience Plan (NRRP), Mission 4 Component 2 Investment 1.3 - Call for tender No. 341 of 15/03/2022 of Italian Ministry of University and Research funded by the European Union – NextGenerationEU, award number PE0000023, Concession Decree No. 1564 of 11/10/2022 adopted by the Italian Ministry of University and Research, CUP D93C22000940001, Project title "National Quantum Science and Technology Institute" (NQSTI). \\
The authors also acknowledge support from Fondazione Cariparma, from Novo Nordisk foundation grant NNF21OC0070832 in the call "Exploratory Interdisciplinary Synergy Programme 2021", from grants PID2020-115221GB-C41/AEI/10.13039/501100011033, TED2021-131447B-C21 and TED2021-131447B-C22 funded by the Spanish MCIN/AEI/10.13039/501100011033 and by the EU “NextGenerationEU”/PRTR, from the Gobierno de Arag\'on (grant E09-17R Q-MAD) and from the CSIC Quantum Technologies Platform PTI-001.\\
A.G.L. acknowledges support from the European Union’s Horizon 2020 research and innovation program under Grant Agreement No. 899354 (SuperQuLAN), and from CSIC Interdisciplinary Thematic Platform (PTI+) on Quantum Technologies (PTI-QTEP+).

\end{document}